\documentclass[a4paper,11pt]{article}

\usepackage{amssymb}
\usepackage[rightcaption]{sidecap}
\usepackage{wrapfig}
\usepackage{multicol}
\usepackage{xcolor}
\definecolor{sectiongrey}{gray}{0.2}
\definecolor{jkasgrey}{gray}{0.8}

\usepackage{booktabs}
\let\oldtabular\tabular
\renewcommand{\tabular}{\small\oldtabular}

\usepackage{abstract}
\makeatletter
\renewcommand*\l@section{\@dottedtocline{1}{0em}{2em}}
\renewcommand*\l@subsection{\@dottedtocline{2}{2em}{3em}}
\renewcommand*\l@subsubsection{\@dottedtocline{3}{5em}{4em}}
\makeatother
\let\savenumberline\numberline
\def\numberline#1{\savenumberline{#1.}}

\renewenvironment{enumerate}%
  {\begin{list}{\arabic{enumi}.}%
     {\topsep=0.5em\itemsep=0.5em\parsep=0mm\partopsep=0mm\usecounter{enumi}}%
   }{\end{list}}

  {\begin{list}{$\bullet$}%
     {\topsep=0.5em\itemsep=0.5em\parsep=0mm\partopsep=0mm}%
   }{\end{list}}

\usepackage{url}
\usepackage{graphicx}
\usepackage[varg]{txfonts}
\usepackage[font=small,format=plain,labelsep=period,labelfont=bf]{caption}

\usepackage{titling}
\pretitle{\begin{center}\LARGE\fontfamily{ptm}\selectfont\bfseries}
\posttitle{\end{center}}

\usepackage{natbib}
\bibpunct{(}{)}{;}{a}{}{,}

\usepackage{geometry}
\usepackage{authblk}

\usepackage{titlesec}
\titlelabel{\thetitle.~~}
\titleformat*{\section}{\fontfamily{ptm}\selectfont\large\centering\bfseries}
\titleformat*{\subsection}{\fontfamily{ptm}\selectfont\centering\bfseries}
\titleformat*{\subsubsection}{\fontfamily{ptm}\selectfont\centering\itshape}
\titlespacing*{\section}{0pt}{*4}{*2}
\titlespacing*{\subsection}{0pt}{*4}{*2}
\titlespacing*{\subsubsection}{0pt}{*2}{*1.2}
\titlespacing*{\paragraph}{0pt}{*1.5}{*1.5}

\renewcommand{\footnoterule}{%
  \kern -3pt
  \textcolor{jkasgrey}{\hrule width 1\textwidth height 1pt}
  \kern 2pt
}
\makeatletter
\renewcommand\@makefntext[1]{\leftskip=0.9em\hskip-0.6em\@makefnmark#1}
\makeatother
\usepackage{setspace}

\usepackage{lastpage}
\usepackage{fancyhdr}
\usepackage[colorlinks,
            pdftex,
            breaklinks,
            linkcolor=blue,
            urlcolor=blue,
            anchorcolor=blue,
            menucolor=blue,
            citecolor=blue]{hyperref}

\title{\vspace{-1em}The {\scshape Capella} Program: Toward A Space-only High-frequency Radio VLBI Network Formed by Small Satellites in Low Earth Orbits}

\date{\small Version 2.0 --- \today}
\author[1,$\star$]{Sascha~Trippe}
\author[2]{Taehyun~Jung}
\author[2]{Jung-Won~Lee}
\author[3]{Jan~Wagner}
\author[2,4]{Jeong-Yeol~Han}
\author[1]{Doohyon~Baek}
\author[5]{Wonseok~Kang}
\author[6]{Jae-Hyun~Kyeong}
\author[7]{Junghwan~Oh}
\author[8]{Jae-Young~Kim}
\author[9]{Jongho~Park}
\author[2]{Sang-Sung~Lee}
\author[10]{Jeffrey~A.~Hodgson}
\author[8]{Taeho~Kang}

\affil[1]{Dept. of Physics \& Astronomy, Seoul National University, Gwanak-gu, Seoul 08826, Korea}
\affil[2]{Korea Astronomy and Space Science Institute, Yuseong-gu, Daejeon 34055, Korea}
\affil[3]{Max Planck Institute for Radio Astronomy, 53121 Bonn, Germany}
\affil[4]{University of Science and Technology, Yuseong-gu, 34113 Daejeon, Korea}
\affil[5]{SpaceBeam Inc., Heungdok-gu, Cheongju 28165, Korea}
\affil[6]{LeO Space Inc., Hanbat National University ICO\#422, Yuseong-gu, 34014 Daejeon, Korea}
\affil[7]{Joint Institute for VLBI ERIC, 7991 PD Dwingeloo, The Netherlands}
\affil[8]{Ulsan National Institute of Science and Technology, Ulju-gun, Ulsan 44919, Korea}
\affil[9]{Dept. of Astronomy \& Space Science, Kyung Hee University, Giheung-gu, Yongin 17104, Korea}
\affil[10]{Dept. of Physics \& Astronomy, Sejong University, Guangjin-gu, Seoul 05006, Korea}
\affil[$\star$]{E-mail: \texttt{trippe@snu.ac.kr}}

\begin{document}
\maketitle

\thispagestyle{fancy}

%%%%%%%%%%%%%%%%%%%%%%%%%%%%%%%%%%%%%%%%%%%%%%%%%%%%%%%%%%%

\begin{abstract}
\noindent\normalsize %\textbf{Abstract:}
Very long baseline radio interferometry (VLBI) with ground-based observatories is limited by the size of Earth, the geographic distribution of antennas, and the transparency of the atmosphere. In this whitepaper, we present the tentative design for a space-to-space VLBI program composed of two missions: \textsc{Mimosa}, a pathfinder, and \textsc{Capella}, a science-grade VLBI observatory.

\textsc{Mimosa} is a two-element space-to-space radio interferometer composed of two small ($\lesssim$250~kg) satellites on co-planar polar circular low Earth orbits. Using single-band, single-circular polarization heterodyne HEMT receivers operating at frequencies around 100~GHz, the interferometer is able to achieve a near-perfect visibility plane coverage and an angular resolution of approximately $35~\mu$as along lines of sight approximately perpendicular to the orbit plane. With an effective collecting area of 0.6~m$^2$ and an instantaneous bandwidth of 4~GHz, one finds a $1\sigma$ total point source sensitivity as good as about 5~mJy (for an on-source time of 250,000 seconds). Making realistic assumptions, the science payload of each satellite has a mass of about 100~kg and consumes about 300~W of power.

\textsc{Capella} comprises four small ($\lesssim$500~kg) satellites in two orthogonal polar low-Earth orbit planes. With single-band heterodyne receivers operating at frequencies around 690~GHz, the interferometer is able to achieve angular resolutions of approximately $7~\mu$as. Within a total observing time of three days, a near-complete \emph{uv} plane coverage can be reached. Given the instantaneous bandwidth of 4~GHz, one finds a $1\sigma$ total point source sensitivity as good as about 1.4~mJy (for an on-source time of 250,000 seconds). We expect the science payload of each satellite to have a mass of about 190~kg and to consume about 380~W of power.

Downlink data rates up to about 100~Gbps can be reached through near-infrared laser communication. The technology for all key components required -- radio telescope, receiver, sampler, recorder, frequency standard, positioning system, data downlink, and pointing control system -- is already available, partially off-the-shelf. The data from the telescopes can be correlated on the ground using dedicated versions of existing Fourier transform (FX) software correlators; in addition to the steps required by VLBI data correlation and calibration in general, dedicated routines will be needed to handle the effects of orbital motion, including relativistic corrections.

With the specifications assumed in this whitepaper, \textsc{Capella} will be able to address a range of science cases, including: the shadows of supermassive black holes; the acceleration and collimation zones of plasma jets emitted from the vicinity of supermassive black holes; the chemical composition of accretion flows into active galactic nuclei through observations of molecular absorption lines; mapping supermassive binary black holes; the magnetic activity of stars; and nova eruptions of symbiotic binary stars -- and, like any substantially new observing technique, has the potential for unexpected discoveries.
\vspace{1\baselineskip}
\end{abstract}

%%%%%%%%%%%%%%%%%%%%%%%%%%%%%%%%%%%%%%%%%%%%%%%%%%%%%%%%%%%

%\newpage 
\tableofcontents 
\newpage

%%%%%%%%%%%%%%%%%%%%%%%%%%%%%%%%%%%%%%%%%%%%%%%%%%%%%%%%%%%

\section{Introduction \label{sec:intro}}

Various astrophysical processes show observational signatures at radio frequencies and on angular scales well below one arcsecond. Active galactic nuclei emit relativistic plasma jets that originate in the vicinity (within few Schwarzschild radii) of supermassive black holes \citep[e.g.,][]{hada2013}. The photon rings around black holes caused by gravitational lensing have angular diameters of tens of microarcseconds  at most \citep[e.g.,][]{ehtc2019}. The spatial distribution and kinematics of methanol masers traces the physical conditions within star forming regions on scales of few astronomical units \citep[e.g.,][]{matsumoto2014}. Last but not least, molecular line absorption on sub-parsec scales provides information on the interaction of inflows, outflows, and turbulence in active galactic nuclei (AGN) \citep[e.g.,][]{sawada2019}.

The well-known relationship $\theta \approx \lambda/B$ between resolution angle $\theta$, wavelength $\lambda$, and aperture (or baseline length) $B$ dictates that radio astronomical observations on angular scales of milliarcseconds and below are only possible by means of very long baseline interferometry (VLBI) \citep[see, e.g.,][for an exhaustive review]{thompson2017}. The Event Horizon Telescope (EHT), a global network of telescopes observing at 230~GHz, reaches angular resolutions of 25~$\mu$as \citep{ehtc2019}. On the one hand, this is an impressive achievement. On the other hand, the EHT illustrates the limitations of ground-based VLBI. Observations at such high frequencies are, if at all, only possible from a limited number of locations around the world, due to the absorption of radio light by atmospheric water vapor; at frequencies $\gtrsim$400~GHz the atmosphere is virtually intransparent everywhere. The limitations imposed by geography lead to a sparse sampling of the $uv$ plane (visibility plane, spatial Fourier plane), implying a low image quality. Indeed, the landmark observations of the black holes M~87* \citep{ehtc2019} and Sgr~A* \citep{ehtc2022} by the EHT required several years of data analysis and calibration and resulted in maps that are known to miss source components on various spatial scales, especially spatially extended emission.

Placing radio antennas in orbits around Earth (space-VLBI) can circumvent some or all of the limitations that affect ground-based VLBI (see, e.g., \citealt{gurvits2020} for a historical review of the concept). The first successful proof-of-concept was provided by the Tracking and Data Relay Satellite System (TDRSS) which combined a communications satellite with two ground stations \citep{levy1986}. The VLBI Space Observatory Programme (VSOP) combined a 10-meter radio telescope on a satellite with several radio observatories on the ground \citep{hirabashi1998}. Likewise, the RadioAstron/Spektr-R mission combined a radio observatory on a high elliptical orbit around Earth with various ground stations \citep[e.g.,][]{smirnova2014}; a successor mission, Millimetron, plans to place a radio observatory at the Sun--Earth Lagrange point L2 \citep[e.g.,][]{likachev2022}. The Black Hole Explorer (BHEX) concept envisions placing a radio observatory capable of observations at frequencies up to about 350~GHz on a circular medium Earth orbit \citep{johnson2024}. Observations of M~87 with a VLBI array that included RadioAstron demonstrated \citep{kim2023} the power of space-VLBI, reaching angular resolutions of $\approx$20~$\mu$as and detecting unexpectedly high ($\gtrsim$10$^{12}$~K) brightness temperatures in the nucleus. However, since they relied on ground stations, those networks were not able to overcome the limitations by the atmosphere. With only one antenna based in space, $uv$ plane coverages were sparse and resulted in elongated elliptical beams (point spread functions) with axis ratios as extreme as 10-to-1 \citep[see also, e.g.,][]{giovannini2018}. As already noted by, e.g., \citet{roelofs2019}, a network of radio observatories exclusively based in space is, in principle, able to achieve excellent angular resolutions and image qualities.

In this whitepaper, we outline the \textsc{Capella} Program, a development program comprising two consecutive space missions. The actual astronomical interferometer is \textsc{Capella} (named, of course, after the quadruple star $\alpha$ Aurigae), a space-only VLBI network composed of four small satellites on low-Earth orbits, each carrying a small radio telescope. A combination of high radio frequencies (690~GHz), sufficient observing times (several days), and various satellite orbits provides unprecedented angular resolutions as good as approximately 7~$\mu$as. However, even though all necessary technologies are available in principle, multi-station space-to-space VLBI has never been attempted before and requires a custom design and construction of most payload components as well as a careful design of mission profiles. Given the lack of relevant experience, it appears advisable to precede the construction of \textsc{Capella}, or any other astronomical space-to-space VLBI network, with a small-scale pathfinder interometer. Therefore, we additionally propose a pathfinder mission, \textsc{Mimosa} (named after the prominent binary star $\beta$ Crucis), a space-to-space two-element interferomer composed of two small satellites on low-Earth orbits. \textsc{Mimosa} is designed to observe a limited number of bright targets at frequencies around 100~GHz (radio W-band). By doing so, \textsc{Mimosa} is going to thoroughly test the entire signal processing chain required for space-to-space VLBI: collecting radio light with a telescope, receiving the radiation, digitizing and storing the signal, and transmitting the data to the ground for processing. Despite its limitations, \textsc{Mimosa} would almost immediately obtain the best-ever interferometric images for a number of astronomical objects.

\section{Reference Interferometers \label{sec:interferometer}}

For the remainder of this whitepaper, we will assume reference configurations for the two interferometers as discussed in the following.

\subsection{\textsc{Capella} \label{sec:ref_capella}}

%%%%%%%%%%%%%%%%%%%%%%%%%%%%%%%%%%%%%%%%%%%%%%%%%%%%%%%%%%%
\begin{figure}[t!]
\centering
\includegraphics[width=72mm]{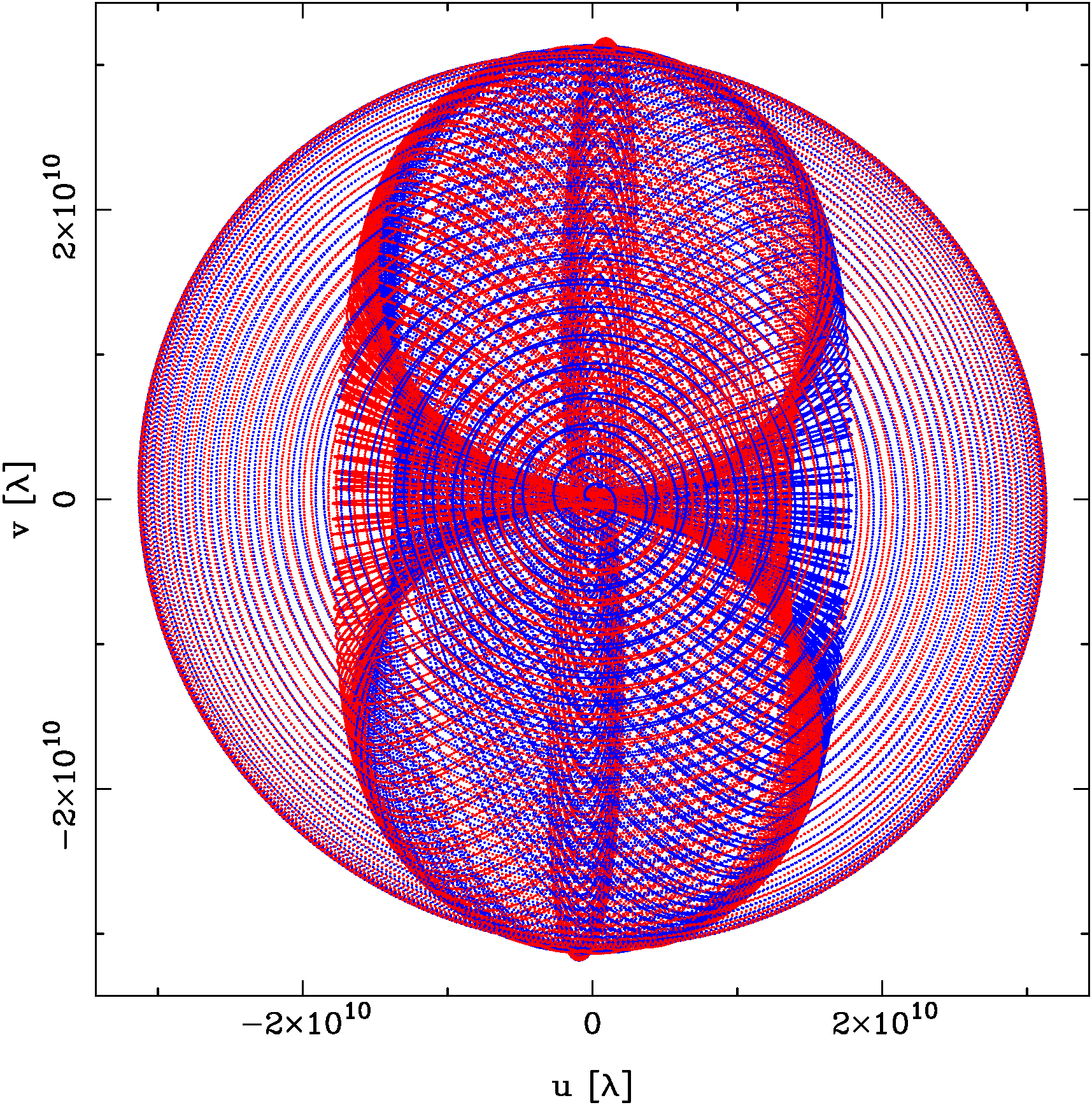}
\hspace{4mm}
\includegraphics[width=72mm]{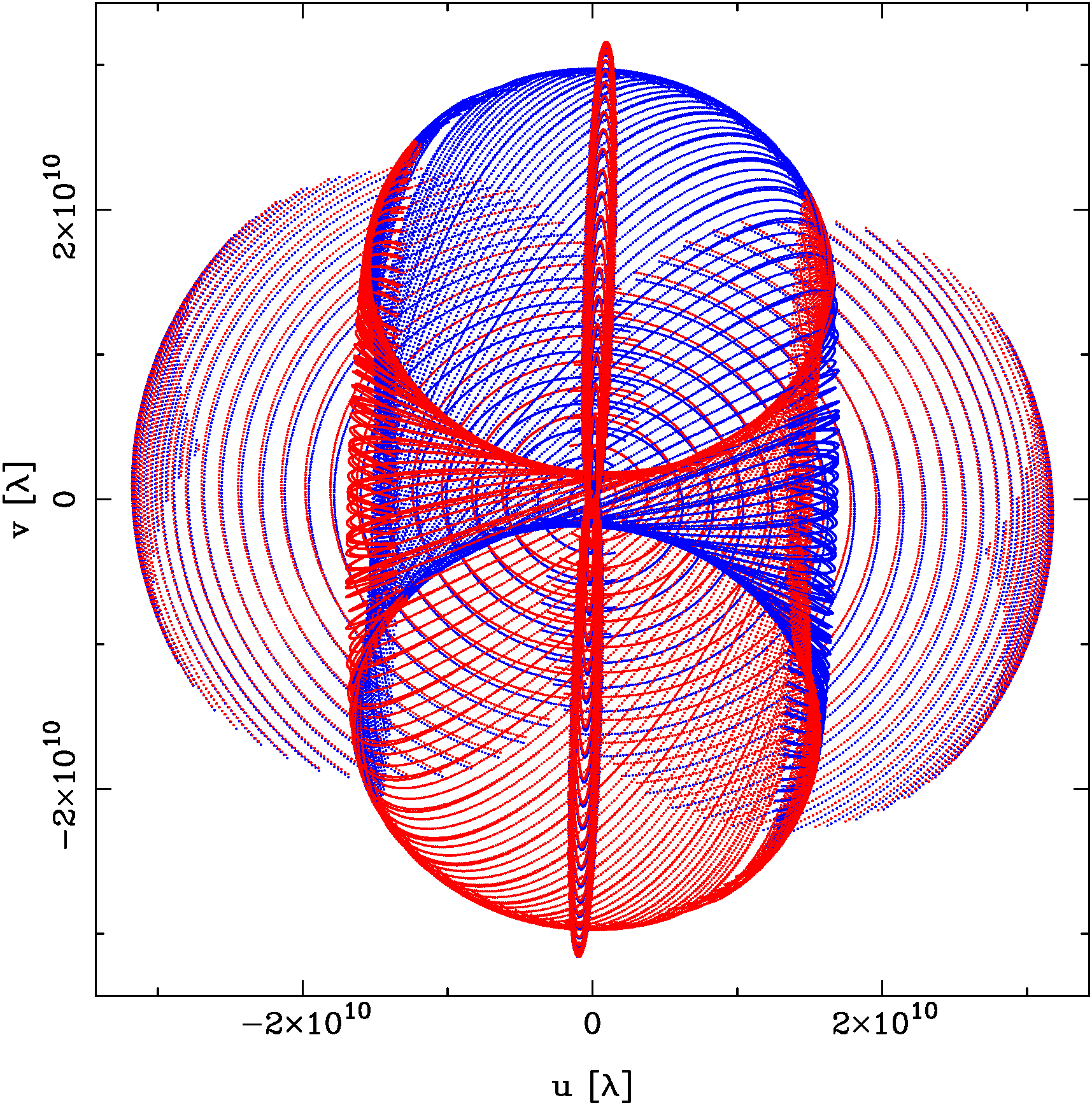} \\[4mm] 
\includegraphics[trim=0mm 0mm 0mm 1mm, clip, width=72mm]{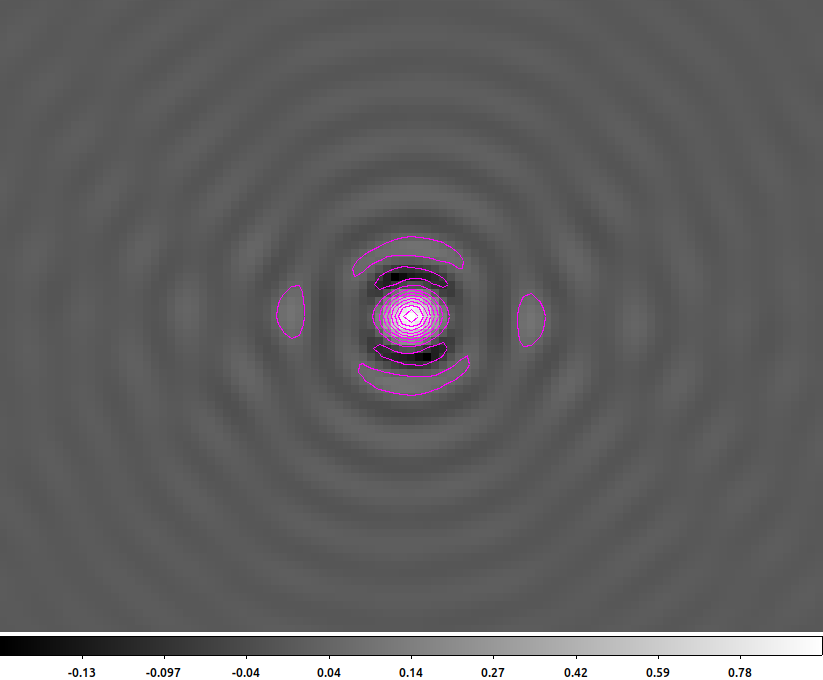}
\hspace{4mm}
\includegraphics[width=72mm]{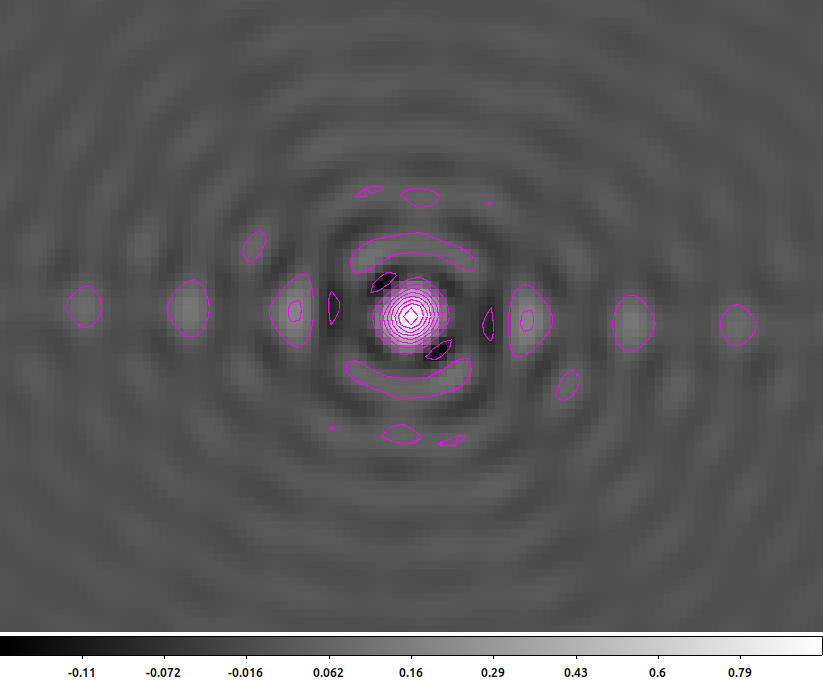}
\caption{\emph{Top panels:} Simulated $uv$ plane coverage achieved by the reference \textsc{Capella} constellation after 250,000 seconds of observing time, assuming observations of M~87 (left) and Sgr~A* (right) at a frequency of 690~GHz, respectively. \emph{Bottom panels:} The corresponding dirty beams, for M~87 (left) and Sgr~A* (right), respectively. The pixel scale is 1~$\mu$as/pixel. The beams are normalized to a maximum value of 1; contour levels are 1, 0.88, 0.76, ..., $-0.08$. The central maxima both have a FWHM of approximately 7~$\mu$as.}
\label{fig:beams_capella}
\end{figure}
%%%%%%%%%%%%%%%%%%%%%%%%%%%%%%%%%%%%%%%%%%%%%%%%%%%%%%%%%%%

%%%%%%%%%%%%%%%%%%%%%%%%%%%%%%%%%%%%%%%%%%%%%%%%%%%%%%%%%%%
\begin{figure}[t!]
\centering
\includegraphics[width=72mm]{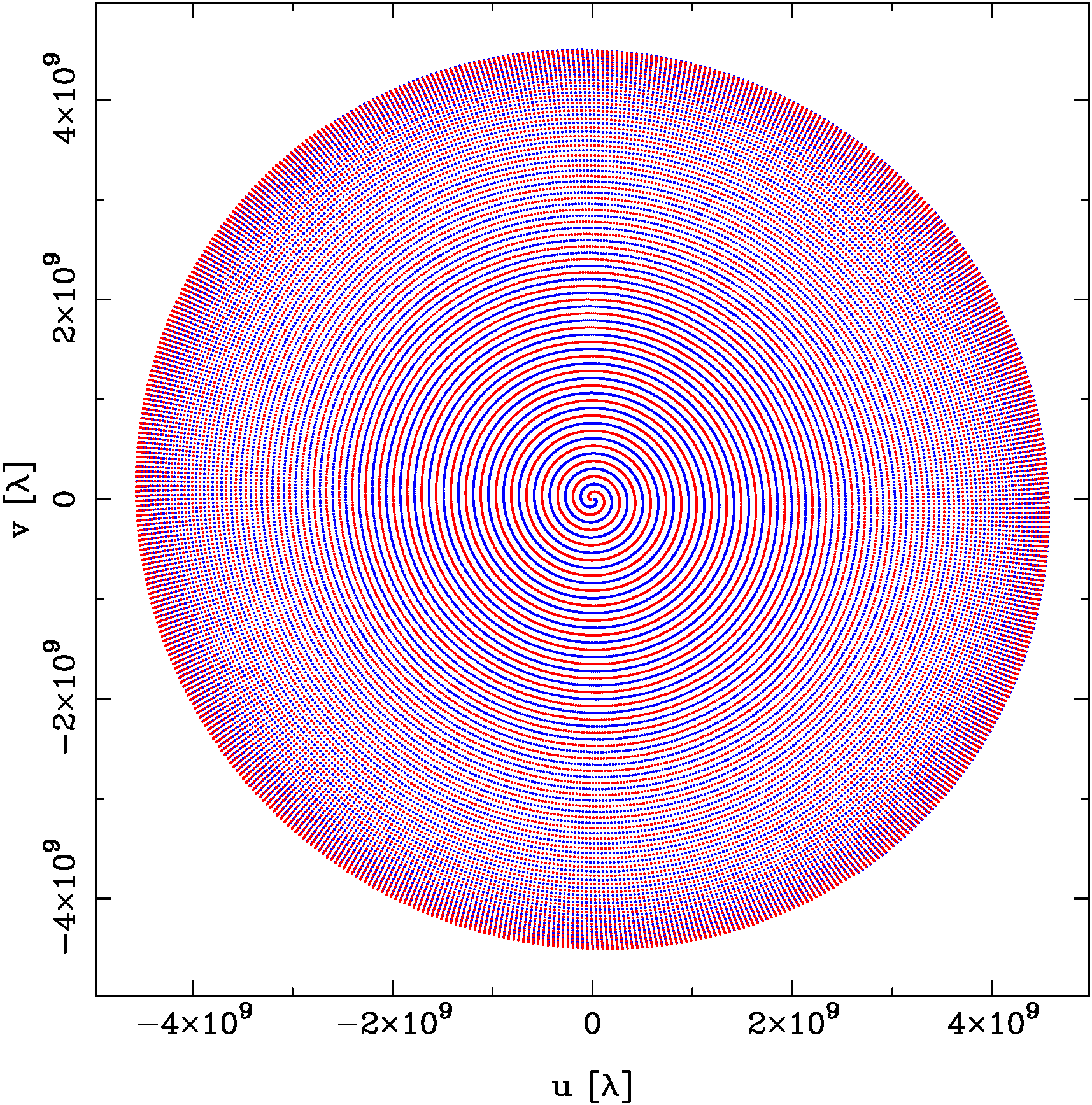}
\hspace{4mm}
\includegraphics[width=72mm]{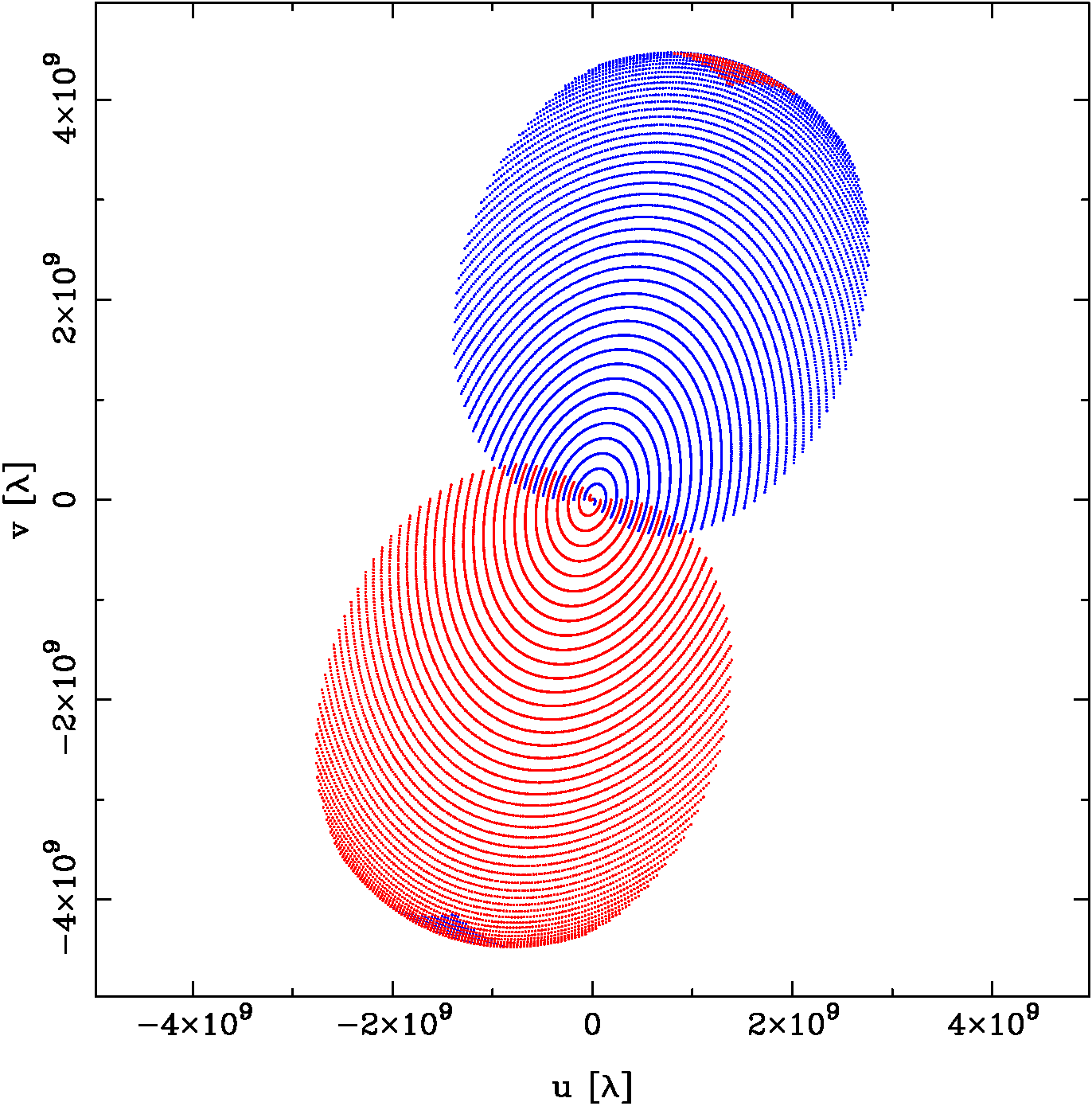} \\[4mm] 
\includegraphics[width=72mm]{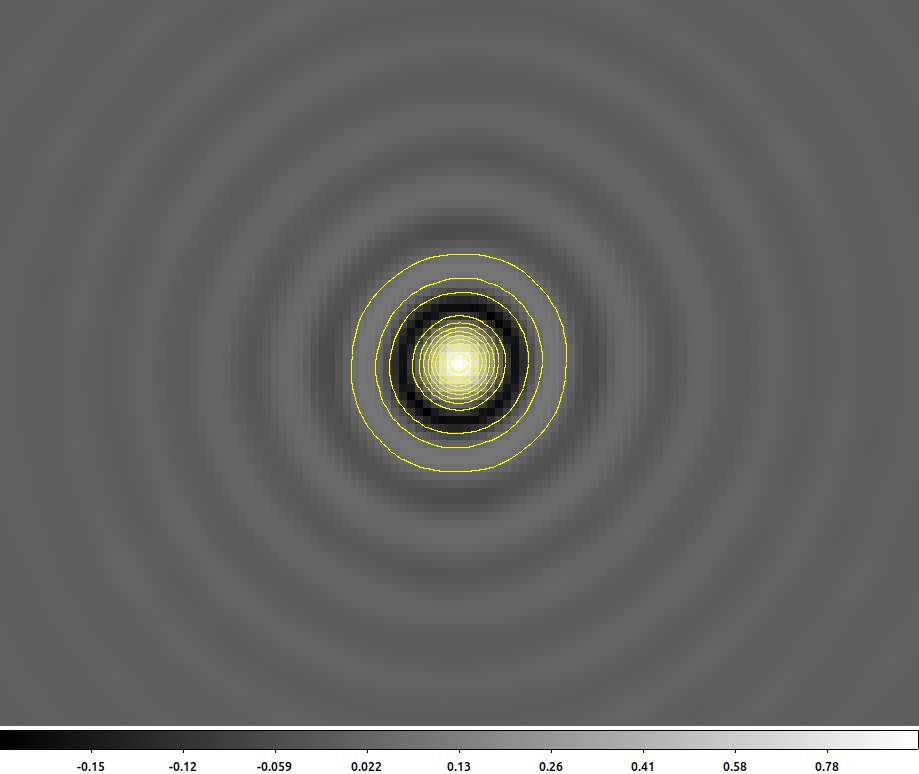}
\hspace{4mm}
\includegraphics[width=72mm]{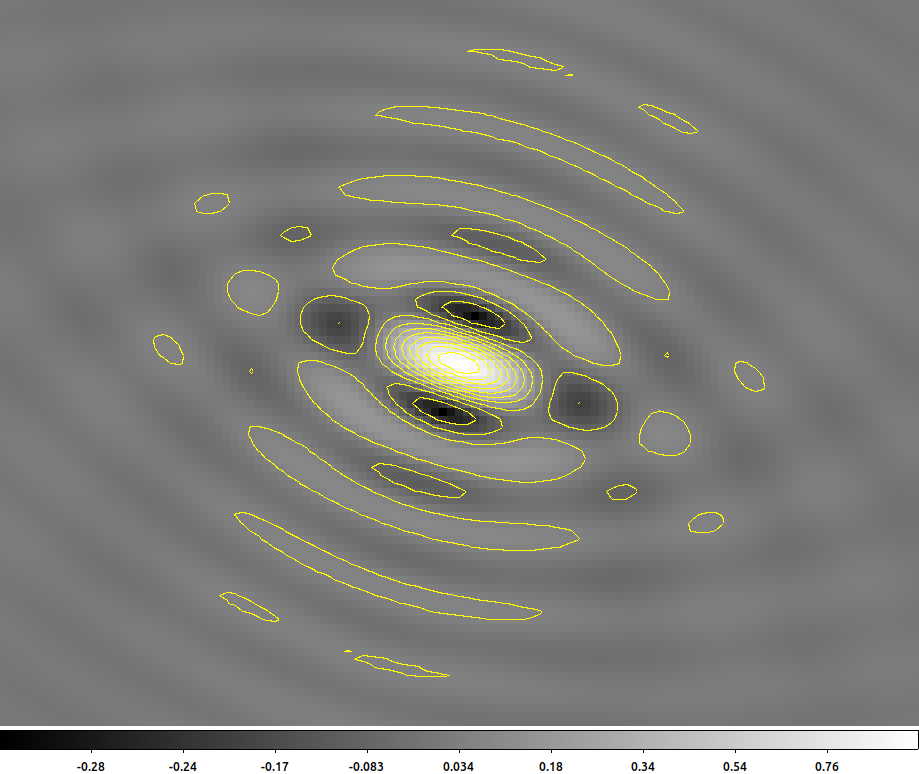}
\caption{\emph{Top panels:} Simulated $uv$ plane coverage achieved by the reference \textsc{Mimosa} constellation after 250,000 seconds of observing time, assuming observations of M~87 (left) and OJ~287 (right) at a frequency of 100~GHz, respectively. Since the line of sight to M~87 is almost perpendicular to the \textsc{Mimosa} orbit plane, the $uv$ coverage is essentially perfect. In the case of OJ~287, the $uv$ tracks are cropped by shadowing by Earth. \emph{Bottom panels:} The corresponding dirty beams, for M~87 (left) and OJ~287 (right), respectively. The pixel scale is 5~$\mu$as/pixel. The beams are normalized to a maximum value of 1; contour levels are 1, 0.88, 0.76, ..., $-0.20$. The central maxima have a FWHM of approximately 35~$\mu$as (circular) for M~87 and 65$\times$35~$\mu$as (elliptical) for OJ~287, respectively.}
\label{fig:beams_mimosa}
\end{figure}
%%%%%%%%%%%%%%%%%%%%%%%%%%%%%%%%%%%%%%%%%%%%%%%%%%%%%%%%%%%

\textsc{Capella} is a constellation of four satellites on four individual low-Earth orbits (LEOs). All orbits are circular (with orbital speeds being the corresponding Newtonian circular speeds) and polar for simplicity; thus we do not need to take into account regression of nodes or rotation of apsides \citep[e.g.,][]{brown1998}. In geocentric equatorial Cartesian coordinates $(x,y,z)$, two satellite orbits are located in the $xz$ plane at altitudes above ground of 500~km and 600~km, respectively; the other two orbits are located in the $yz$ plane at altitudes of 450~km and 550~km, respectively. Observations are conducted at a frequency $\nu$ of 690~GHz (i.e., a wavelength of 434~$\mu$m) with an instantaneous bandwidth $\Delta\nu$ of 4096~MHz. The effective collecting area $A_{\rm e}$ is 2.4~m$^2$, implying an antenna efficiency $\rho_{\rm A} = 2 k_{\rm B} / A_{\rm e}$ (with $k_{\rm B}$ being Boltzmann's constant) of 1150~Jy\,K$^{-1}$. Each satellite carries a single-circular polarization double-sideband (DSB) receiver; the DSB system temperature $T_{\rm sys}$ is 100~K.

As usual for space-based astronomical missions, observing runs can, at least in principle, be of arbitrary duration; observing times of X-ray space observatories are commonly given in megaseconds. The $uv$ plane coverage for a total observing time of 250,000 seconds, along with the resulting dirty beams (point spread functions), is illustrated in Figure~\ref{fig:beams_capella} for two targets: the active galaxy M~87 with J2000 equatorial coordinates $(\alpha,\delta) = (187.7^{\circ}, 12.4^{\circ})$; and the black hole in the center of the Milky Way, Sagittarius~A* (Sgr~A*) with $(\alpha,\delta) = (266.4^{\circ}, -29.0^{\circ})$. Satellite positions were calculated in time increments of 10 seconds. At each time step, all satellite positions were projected onto a plane P through $(x,y,z)=(0,0,0)$ perpendicular to the line of sight using Rodrigues' formula; the plane was then rotated such that the pairwise distance vectors $(u,v)$ in the plane agreed with the convention for $uv$ coordinates (i.e., $u$ corresponding to north--south distances and $v$ corresponding to east--west distances, in units of the observing wavelength). For a given $uv$ plane distribution, the map of the dirty beam, in angular coordinates $(l,m)$, followed from assigning a value of 1 to points falling onto $uv$ tracks and a value of 0 elsewhere (``uniform weighting''), and computing a two-dimensional discrete Fourier transform. The extension and pixel scale of the $uv$ map were adjusted such that the pixel scale of the $lm$ plane map was 1~$\mu$as/pixel.

Like any space telescope on a LEO, the \textsc{Capella} satellites are shadowed (from the point of view of the target) by Earth for parts of their orbits. We considered a satellite shadowed if it was located so far behind the plane P that the angle satellite--center of Earth--P exceeded 15$^{\circ}$; for the lowest of our four circular orbits, this implies a minimum allowed distance between the line of sight and the surface of Earth of 220~km and a maximum distance behind P of about 1770~km.

\subsection{\textsc{Mimosa} \label{sec:ref_mimosa}}

\textsc{Mimosa} is formed by two satellites on two individual low-Earth orbits (LEOs). Both orbits are circular and polar. Furthermore, we assume the orbits to be coplanar, since this requires only a single launch. The two satellite orbits are located in the $yz$ plane at altitudes above ground of 550~km and 600~km, respectively. Observations are conducted at a frequency $\nu$ of 100~GHz (i.e., a wavelength of 3~mm) with an instantaneous bandwidth $\Delta\nu$ of 4096~MHz. The effective collecting area is 0.6~m$^2$, implying an antenna efficiency of 4600~Jy\,K$^{-1}$. Each satellite carries a single-circular polarization single-sideband (SSB) receiver; the system temperature $T_{\rm sys}$ is 40~K.

The $uv$ plane coverage for a total observing time of 250,000 seconds, along with the resulting dirty beams (point spread functions), is illustrated in Figure~\ref{fig:beams_mimosa} for two targets: the active galaxy M~87 with J2000 equatorial coordinates $(\alpha,\delta) = (187.7^{\circ}, 12.4^{\circ})$; and the blazar OJ~287 with $(\alpha,\delta) = (133.7^{\circ}, 20.1^{\circ})$. Satellite positions were calculated in time increments of 10 seconds. We considered a satellite shadowed if it was located so far behind the plane P that the angle satellite--center of Earth--P exceeded 15$^{\circ}$; for the lower of our two circular orbits, this implies a minimum allowed distance between the line of sight and the surface of Earth of 310~km and a maximum distance behind P of about 1790~km. The extension and pixel scale of the $uv$ map were adjusted such that the pixel scale of the $lm$ plane map was 5~$\mu$as/pixel.

\section{Science Cases \label{sec:science}}

\textsc{Capella} will be able to observe astrophysical objects which (1) are sufficiently bright at its observing frequency, with flux densities of at least tens of milli-Janskys; and (2) whose emission is sufficiently spatially concentrated, meaning angular scales between a few $\mu$as and a few mas. These constraints naturally suggest the following, not exhaustive, list of science cases.

\subsection{Processes Involving Supermassive Black Holes \label{sec:blackholes}}

\subsubsection{Black Hole Shadows \label{sec:shadows}}

The general relativistic description of black holes implies the existence of a distinct photon orbit with a radius of three gravitational radii for non-rotating black holes and smaller radii for rotating black holes, assuming prograde rotation in the equatorial plane \citep{bardeen1972}. Due to gravitational lensing by the black hole, the photon ring, and thus the region  around the black hole which is void of light (``black hole shadow''), appears under a radius of $\sqrt{27}$ gravitational radii to an observer at infinity, regardless of the black hole spin \citep{falcke2000}. The gravitational radius is given by $R_{\rm g} = G M_{\bullet} / c^2$, with $M_{\bullet}$ being the black hole mass, $G$ denoting Newton's constant, and $c$ being the speed of light. Sufficient equipment provided, an observer at a (non-cosmological) distance $D$ is able to image a ring with an angular radius of $\xi = \sqrt{27} R_{\rm g} / D$. If the distance $D$ is known (like in the case of M~87*), it is possible to derive the black hole mass from $\xi$. If the ratio of the mass of and the distance to the black hole is known (like in the case of Sgr~A*), it is possible to compare the observed image to the theoretically expected one and to search for deviations from general relativity \citep{ehtc2022b}.

The photon rings of both M~87* and Sgr~A* have angular diameters of about 40~$\mu$as and 50~$\mu$as, respectively, and could be resolved by the EHT. Given that the angular resolution of \textsc{Capella}, $\approx$7~$\mu$as, is about four times better than that of the EHT, it will be able to spatially resolve the photon rings of black holes with mass-to-distance ratios about four times smaller than those of M~87* or Sgr~A*. The central black hole in the giant elliptical galaxy IC~1459, located at a distance of about 30~Mpc, has a mass of about $2.6\times10^9\,M_\odot$ \citep{cappellari2002}; the diameter of the photon ring (i.e., $2\xi$) is thus about 9~$\mu$as. The mass of the black hole in the giant elliptical galaxy M~84, located 17~Mpc away, is estimated to be about $8.5\times10^8\,M_\odot$ \citep{walsh2010}, corresponding to $2\xi \approx 5~\mu$as. The black hole in the Andromeda galaxy, M~31*, located 760~kpc from the Milky Way, has a mass of about $2.6\times10^9\,M_\odot$ (and thus $2\xi \approx 19~\mu$as; \citealt{bender2005}); its radio emission is usually too faint ($<$1~mJy) for a \textsc{Capella} type observatory but might reach sufficient levels during phases of higher-than-average accretion rates. Theoretical population studies of supermassive black holes show that a  \textsc{Capella} type interferometer should be able to resolve the shadows of about ten black holes \citep{pesce2021}.

Notably, the masses of supermassive black holes are usually not well constrained even in the local universe; a case in point is the black hole in M~84 for which masses as low as $4\times10^8\,M_\odot$ (corresponding to $2\xi \approx 2.4~\mu$as) and as high as $1.5\times10^9\,M_\odot$ (meaning $2\xi \approx 9~\mu$as) have been proposed \citep[see][and references therein]{walsh2010}. Therefore, a systematic survey of the centers of active galaxies in the local universe with \textsc{Capella} will be able to either measure directly (if the shadow is resolved) or place upper limits on (if the shadow is unresolved) the masses of their central supermassive black holes.

\textsc{Capella} will be able to provide both a high angular resolution and, thanks to its dense sampling of the $uv$ plane, high-quality images (cf. Figure~\ref{fig:beams_capella}). This makes it possible to study potential deviations from a ringlike geometry with unprecedented accuracy. A high image quality opens a path toward accurate tests of the Kerr metric (see, e.g., \citealt{bambi2017} for a review). Even when assuming the validity of general relativity (which is a rather safe assumption for most purposes), the observable morphology of photon rings depends on the astrophysical processes occurring around the black holes; depending on whether the light concentrated in the ring originates from an accretion disk or a jet, and also depending on the viewing geometry, images of black hole shadows will show distinct differences \citep[see, e.g.,][]{akiyama2015}. Characteristic asymmetries, and their variation with time, provide information on the internal physics of accretion flows and the spin of the black hole \citep{wielgus2020}.

\subsubsection{Jets of Active Galactic Nuclei \label{sec:agnjets}}

Active galactic nuclei (AGN) are powered by the accretion of interstellar matter into supermassive black holes. A fraction of AGN shows highly collimated bipolar outflows of magnetized plasma moving at (usually) relativistic speeds, \emph{jets}, which reach from the immediate environment of the black hole out to, in extreme cases, hundreds of kiloparsecs; the jet of M~87 is a notable example. The interaction of AGN jets with the interstellar gas in a galaxy is supposed to be a driver of \emph{AGN feedback}, meaning that the resulting heating or expulsion of the gas can suppress star formation and thus control the evolution of that galaxy \citep[see, e.g.,][for a review]{fabian2012}. Jets arise from an interaction of interstellar gas, magnetic fields, and rotation of either the black hole itself \citep{blandford1977} or of the accretion disk around the black hole \citep{blandford1982} (both processes may contribute to the same jet; e.g., \citealt{hardee2007}). Despite their importance, the launch, acceleration, and collimation processes that control AGN jets are only partially understood. While some jets are known to form within 10 Schwarzschild radii from the black hole \citep{hada2013} and various observations provide several constraints, it has not been possible to conclusively pin down a specific formation process. While the EHT observations of M~87 clearly show the photon ring around the black hole, they do not show the foot point of the jet; arguably, this is due to the insufficient \emph{uv} plane coverage provided by the EHT which filters out spatially extended emission.
Due to its dense sampling of the $uv$ plane, \textsc{Capella} will be able to map jets and the inflow/outflow structure around supermassive black holes on all spatial scales.

Our lack of understanding of jet physics is emphasized by the curious case of Sgr~A* which, like M~87*, has been resolved by the EHT. The spectral energy distribution of Sgr~A* indicates the presence of a jet \citep{markoff2007}; however, observations have consistently failed to find one. VLBI observations that combine both high angular resolution and high image quality will be required to conclusively address the question if Sgr~A* emits jets. Since Sgr~A* is a prototypical low-luminosity AGN (LLAGN), with luminosities that are at least eight orders of magnitude below the Eddington limit, its case also raises the question if, and how, jets can form in LLAGN in general. A dedicated \textsc{Capella} survey of nearby LLAGN, mapping structures as small as tens of Schwarzschild radii, will provide new insights and constraints.

AGN jets are emitters of synchrotron radiation, meaning that their opacity is a function of frequency. The radio cores of radio-bright AGN, i.e. the \emph{apparent} starting points of their jets, tend to be located closer to the central black hole with increasing frequency; this \emph{core shift effect} \citep[e.g.,][]{soko2011} implies that many radio cores are the surfaces of unit optical depth within the jet plasma, meaning in turn that observations aimed at the origin of jets need to be performed at high frequencies to overcome the limitations set by opacity. Polarimetric studies of blazar jets indicate that observations at frequencies of several hundred GHz penetrate the jet plasma and actually trace physical structures, like recollimation shocks \citep{park2018}. Accordingly, not only the angular resolution and image quality, but also the high observing frequency of \textsc{Capella} will be needed for distinguishing radio cores and actual starting points of jets, and for gaining a complete picture of AGN jet physics.

\subsubsection{Molecular Gas Flows \label{sec:agnlines}}

The rest-frame frequency range $690 \pm 10$~GHz contains 200 known cosmic molecular radio lines, including, most famously, the CO(6--5) transition, multiple transitions of molecules as complex as CH$_3$CH$_2$CN, and several lines of unknown origin (according to the \href{https://physics.nist.gov/cgi-bin/micro/table5/start.pl}{NIST catalog of interstellar molecular microwave transitions}). Active galactic nuclei are the most energetic persistent sources of radiation in the universe; they emit non-thermal continuum radiation that is spatially highly concentrated (brightness temperatures $\gtrsim$10$^{10}$~K). Interstellar matter in the line of sight toward the AGN can absorb this radiation at the frequencies corresponding to molecular transitions, leading to characteristic absorption line spectra. The various absorption lines provide important information on the composition and temperature (through line strengths and line ratios), kinematics (through the Doppler shift), and structure (lines at different velocities probing different gas layers) of the gas around supermassive black holes \citep[e.g.,][]{rose2020}.

On (sub-)parsec physical scales, corresponding to (sub-)milliarcsecond angular scales for most extragalactic sources, little is known about the distribution and the dynamics of the interstellar gas. On those scales, AGN are commonly assumed to show both inflows and outflows, as well as a circumnuclear gas and dust torus, but very few absorption line observations on VLBI scales have ever been conducted. Observations of the radio galaxy NGC~1052 \citep{sawada2019} have been able to identify a molecular gas torus on scales of around one parsec, to constrain the kinematics of the jet that originates from the black hole, and to constrain the gas density.

Observations with \textsc{Capella} will be able to extend this type of analysis to unprecedentedly small spatial scales. Angular resolutions of $\gtrsim$7~$\mu$as correspond to scales of tens to hundreds of gravitational radii for black holes in nearby galaxies; due to the combination of high spatial resolution and line-of-sight kinematics \textsc{Capella} will thus be able to provide detailed three-dimensional maps of the gas inflows to and outflows from AGN.

\subsubsection{Binary Black Holes \label{sec:bbh}}

Arguably most, if not all, galaxies host supermassive black holes with masses ranging from millions to billions of solar masses in their centers. The paradigm of hierarchical structure formation in the universe, which includes the frequent occurrence of galaxy--galaxy mergers, suggests that at least some galaxies host black hole binary systems. Whereas they are theoretically expected, observational evidence for binary supermassive black holes is very sparse and often ambiguous.

Over the years, various candidates for AGN powered by binary black holes have been suggested, including some radio-bright ones. Arguably the most promising candidate identified as yet is the blazar OJ~287 \citep{pietila1998, dey2021}. Modeling the variability of its long-term ($>$100 years) light curve with a binary black hole model suggests that, to good approximation, the less massive ($\approx10^8~M_\odot$) component orbits the more massive ($\approx1.5\times10^{10}~M_\odot$) one on an elliptical orbit with a semimajor axis length of about 0.052~pc. Given the redshift $z=0.306$, this distance corresponds to an angular separation of about 8~$\mu$as.\footnote{All angular scales quoted in this section are obtained with information from the \href{https://ned.ipac.caltech.edu/}{NASA/IPAC Extragalactic Database} and assume a cosmology with $H_0 = 67.8$~km\,s$^{-1}$\,Mpc$^{-1}$, $\Omega_{\rm m} = 0.308$, and $\Omega_\Lambda = 0.692$.} \citet{lobanov2005} analyzed the structural variations in the parsec-scale jet of the quasar 3C~345 and concluded that these variations agree with orbital motion of a binary black hole with a separation between the components of about 0.33~pc; at $z=0.593$, this corresponds to an angular separation of about 26~$\mu$as. The quasar PG 1302-102 shows periodic optical variability that is best explained by a supermassive binary black hole with a separation of $<$1~pc, corresponding to $<$170~$\mu$as at $z=0.278$ \citep{graham2015}.

For the aforementioned cases, the relevant angular scales are on the order of tens of microarcseconds, source structures are usually complex due to the presence of jets, and expected orbital time scales are on the order of years. These properties make binary black hole candidates obvious targets for systematic time-resolved mapping studies with \textsc{Capella}, ideally resulting in movies of orbiting accreting black holes. Indeed, the recent population study of \citet{zhao2024} concluded that a \textsc{Capella}-like VLBI array should be able to detect about 20 supermassive black hole binaries at redshifts up to 0.5.

\subsection{Processes Involving Stars \label{sec:stars}}

\subsubsection{Flares \label{sec:flares}}

Stars other than our Sun are not usually considered radio sources. However, they show a range of transient phenomena, especially flares, that can emit large amounts of non-thermal radiation at high radio frequencies. The M dwarf stars Proxima Centauri and AU Microscopii are known for radio flares with flux density levels of tens of mJy at 220~GHz occurring multiple times per day \citep{macgregor2020}. Both stars have angular diameters of around or slightly below 1~mas, meaning that \textsc{Capella} will be able to resolve radio bright structures on their surfaces as small as about 1\% of the stellar diameter. Since M dwarf flares, as far as known, rise and decay on time scales of minutes, they cannot be mapped by a \textsc{Capella} type interferometer in a straightforward manner as this requires observing times of at least several hours. Instead, a dedicated monitoring observation lasting multiple days could pinpoint specific hot spots of activity on the stellar surface, provided those exist, by mapping their time-integrated emission. Likewise, a sufficiently bright flare, even if only lasting for minutes, would provide at least a limited set of $uv$ plane data, thus making it possible to at least roughly localize the origin of the emission on the stellar surface.

Flaring due to magnetic activity has also been observed in young stellar objects. \citet{bower2003} observed several millimeter-wavelength flares from a T~Tauri object in the Orion nebula; altogether, the enhanced radio emission lasted for about one week. A solar-sized object located at a distance of 400~pc has an angular diameter of 23~$\mu$as. Accordingly, \textsc{Capella} observations initiated shortly after the onset of enhanced activity would be able to map such flares on or above the stellar surface.

\subsubsection{Symbiotic Stars \label{sec:novae}}

Symbiotic stars are interacting binary stars composed of an evolved red giant and a hot companion star, like a white dwarf. Transfer of matter from the giant star to the compact companion can lead to recurrent nova eruptions. A recent example is RS~Ophiuci which last erupted in 2021. Radio interferometric observations at 5~GHz obtained a few weeks after the eruption show an extended bipolar outflow spanning about 240~au; comparing the outflow geometry to the line-of-sight kinematics derived from optical emission lines made it possible to constrain the geometry of the system \citep{munari2022}. At the distance of RS~Oph ($\approx$2.7~kpc), the angular resolution of \textsc{Capella} ($\approx$7~$\mu$as) corresponds to about four solar radii. Accordingly, mapping Galactic novae with \textsc{Capella} will provide new insights into the physical processes in the circumstellar environment around erupting white dwarfs, especially the ejection of matter and the formation of bipolar outflows.

\subsubsection{Interstellar Near-field VLBI \label{sec:nearfield}}

The design of VLBI observations almost universally assumes that the light from the target can be described as propagating with plane wave fronts. This is indeed the case for emitters located at distances $D \gg D_{\rm F}$ (Fraunhofer condition), with the Fraunhofer distance being
\begin{equation}
\label{eq:fraunhofer}
D_{\rm F} = \frac{2B^2}{\lambda} ~ .
\end{equation}
For sources located within the Fresnel region (near-field region) of a VLBI observatory -- i.e., $D \lesssim D_{\rm F}$ -- the curvature of the incoming wavefront becomes a measurable physical quantity. In this case, the interferometric delay depends explicitly on the finite distance to the source, requiring spherical delay modeling to account for corrections of order $B^{2}/2D$ \citep{Sekido2006}.

For a VLBI network like \textsc{Capella}, with $B\approx14$,000~km and $\lambda=434~\mu$m, the Fraunhofer distance is $D_{\rm F} \approx 29$~pc. This implies that radio flare stars (cf. Section~\ref{sec:flares}) like Proxima Centauri (located at a distance of 1.3~pc) and AU Microscopii (at 9.7~pc) are located within the Fresnel region. Thus, at least in theory, the observatory can probe the three-dimensional structure of their radio emitting regions. For the case of \textsc{Mimosa} ($B\approx14$,000~km, $\lambda=3$~mm), one finds $D_{\rm F} \approx 4$~pc, which is still well beyond the distance to Proxima Centauri.

\section{System Components \label{sec:system}}

\begin{figure}[t!]
\centering
\includegraphics[trim=20mm 30mm 5mm 5mm, clip, width=130mm]{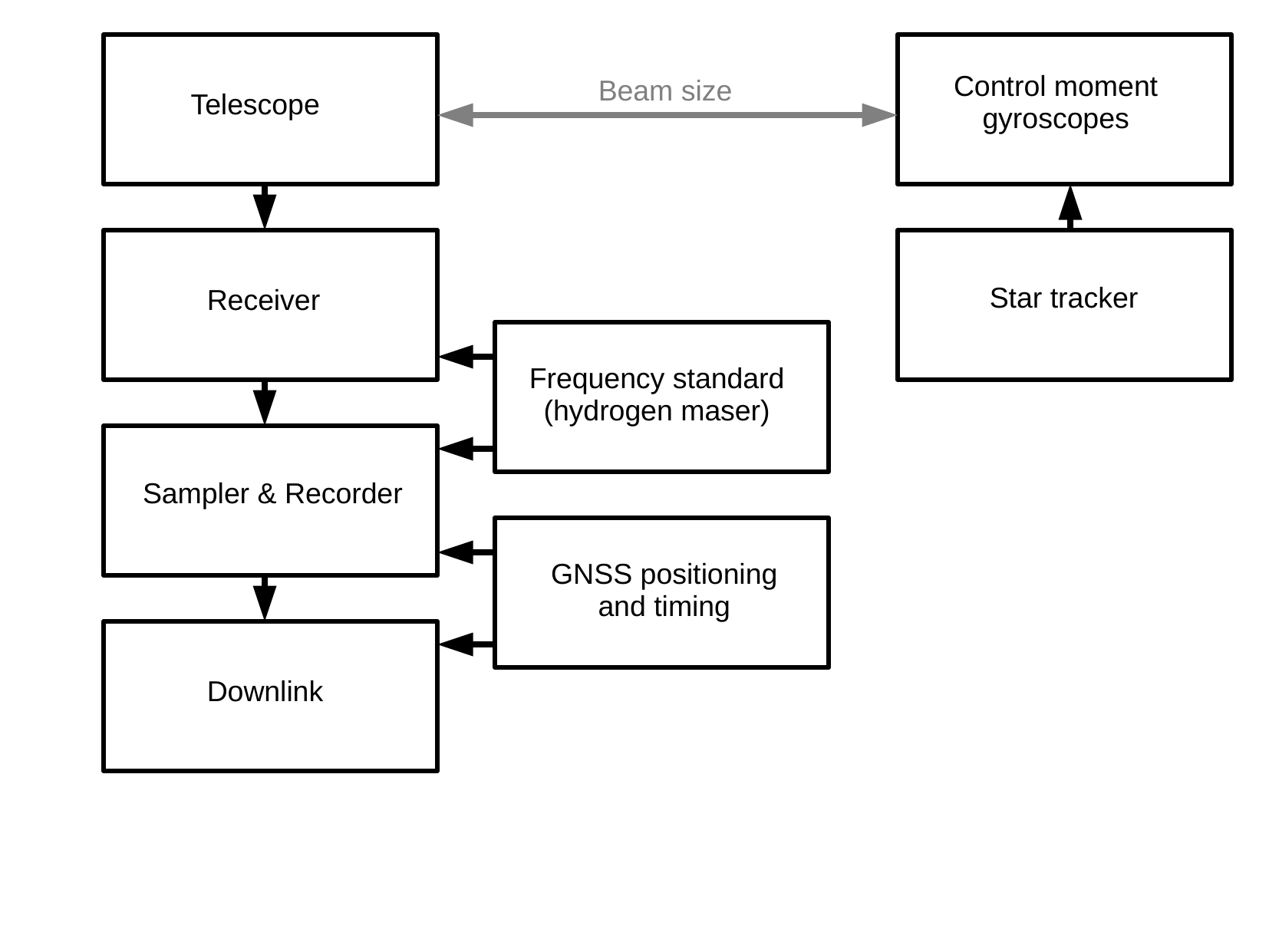}
\caption{The main components of a VLBI satellite, with their mutual interactions (black arrows) and other dependencies (grey arrows).}
\label{fig:system}
\end{figure}

The \textsc{Capella} and \textsc{Mimosa} interferometers are each composed of several space observatories that operate as VLBI stations. Each satellite needs to carry the components required for a VLBI station: a telescope to collect the light; a receiver to convert the radio light to an electric signal; a recording system that digitizes and records the signal; a precision frequency standard; a Global Navigation Satellite System (GNSS) positioning and timing system for accurate positioning and timing; and a fast downlink to transmit the data to a ground station (this may be compared to the layout of a modern ground-based high-frequency VLBI system; e.g., \citealt{ehtc2019b}). In addition, a spacecraft orientation and pointing system needs to be implemented. The necessary components are summarized in Figure~\ref{fig:system} and discussed one-by-one in the following.

\subsection{Platform \label{sec:platform}}

Satellites with launch masses $\lesssim$500~kg and launch diameters $\lesssim$2~m are commonly considered ``small''; satellites in this category are relatively inexpensive and can be mass produced, up to batches of several thousand units for current and upcoming LEO communication satellite constellations. For these reasons, we have designed the \textsc{Capella} and \textsc{Mimosa} systems such that the individual space observatories (VLBI stations) can be accommodated by small satellites as their carrier platforms; examples for commercially available small satellite platforms are the Korea Aerospace Research Institute (KARI) Compact Advanced Satellite 500 / KOMPSAT Next (\href{https://www.kari.re.kr/eng/contents/170}{CAS500}), the Lockheed Martin \href{https://www.lockheedmartin.com/content/dam/lockheed-martin/space/documents/lm400/LM400-LM2100_LM50_Product_Card_Final_Web.pdf}{LM~50} series, and the Airbus \href{https://airbusus.com/wp-content/uploads/2023/08/ARROW-Family-Brochures.pdf}{ARROW} family. The carrier satellite will provide elementary functions like power supply and orbit control. In addition, the onboard radio receiver requires cryogenic cooling (to be discussed in Section~\ref{sec:receiver}); thus, the carrier satellite will need to provide a sunshade to stabilize the temperature of the telescope and the receiver compartment through passive cooling. Such a design is common for infrared space telescopes; a sun shield can be combined with a solar panel assembly, like in case of the \href{https://www.spitzer.caltech.edu/mission/mission-overview}{\textsc{Spitzer}} Space Telescope, or can be designed as a flexible stand-alone structure, like in case of \href{https://spherex.caltech.edu/page/about-the-mission}{SPHEREx}. Furthermore, thermal energy storage by phase-change materials may be considered \citep{elshaer2023}.

\textsc{Mimosa} uses two coplanar polar low-Earth orbits; thus, both satellites can share the same launch vehicle. The \textsc{Capella} constellation uses two orthogonal sets of polar orbits; accordingly, two separate launches will be required, each carrying two satellites. Given the importance of constellations of small satellites for the ``new space age'', multiple launchers have been developed, or are under development, for transporting such satellites into low Earth orbits inexpensively. The Korea Space Launch Vehicle II (\href{https://www.kari.re.kr/eng/contents/176}{KSLV-II}) is able to carry a payload of 1500~kg into a LEO. Very recent cost-effective launcher designs include the Isar Aerospace \href{https://www.isaraerospace.com/spectrum}{Spectrum} (1000~kg payload to LEO), the Rocket Factory Augsburg \href{https://www.rfa.space/rfa-one/}{RFA~1} (1300~kg payload to LEO), the Innospace \href{https://www.innospc.com/page/sub01_03_1}{HANBIT-Mini} (1300~kg payload to LEO), and the Korean Agency for Defense Development's \href{https://english.hani.co.kr/arti/english_edition/e_national/1119166}{sea-launched solid-fuel rocket} (up to 1500~kg payload to LEO). Any launch site that can target polar orbits (like, e.g., the Naro Space Center in case of KSLV-II) is acceptable.

\subsection{Telescope \label{sec:telescope}}

\begin{figure}[t]
\centering
\includegraphics[width=70mm, trim=5mm 10mm 0mm 10mm, clip]{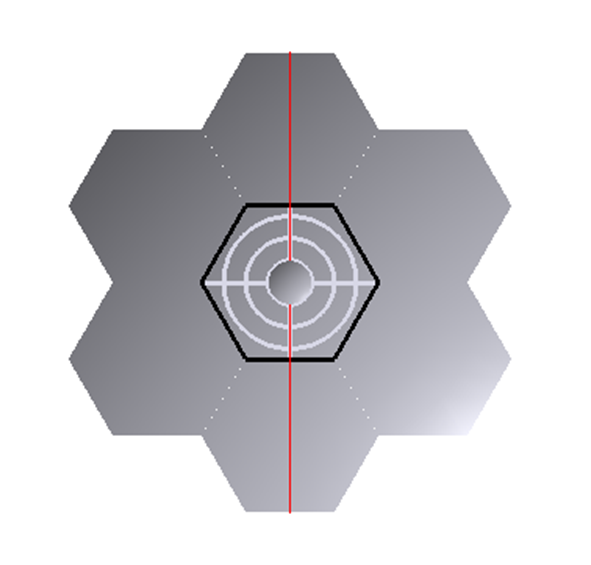}
\includegraphics[width=79mm, trim=0mm 5mm 5mm 10mm, clip]{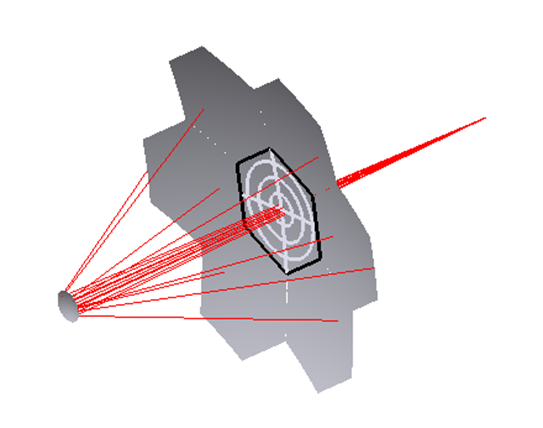}
\caption{Example design for a segmented \textsc{Capella} primary mirror. Each hexagon has a flat-to-flat diameter of 70~cm. \emph{Left:} Front view. \emph{Right:} Angled view, with selected light rays in red.}
\label{fig:mirror}
\end{figure}

The high observing frequency of 690~GHz targeted by \textsc{Capella} requires a telescope with a surface accuracy which is better than that of common radio telescope designs; at the same time, we need to minimize the telescope mass as far as possible. These conditions are met by silicon carbide (SiC) structures. The SiC Cassegrain telescope of the \textsc{Herschel} infrared space telescope had a primary mirror with a diameter of 3.5~m and a thickness of 2.5~mm, and had a total mass of 300~kg. The total r.m.s. wavefront error was $\sigma_{\rm WF} = 6~\mu$m \citep{sein2003}. For the \textsc{Capella} observing wavelength of $\lambda = 434~\mu$m, a wavefront error of 6~$\mu$m corresponds to an aperture surface efficiency of
\begin{equation}
\label{eq:eta_A}
\eta_{\rm A} = \exp[-(4 \pi \sigma_{\rm WF} / \lambda)^2] = 0.97 ~ .
\end{equation}
Assuming a cartridge-type receiver (to be discussed in Section~\ref{sec:receiver}) installed in the Cassegrain focus, the total aperture efficiency, meaning the fraction of the energy arriving at the telescope that is actually captured by the receiver, is approximately 80\% \citep[cf.][]{baryshev2015}.

Using the \textsc{Herschel} telescope data from \citet{sein2003} as a reference, and assuming that (1) the mass of the optical surfaces scales with their surface area (since they cannot be made substantially thinner anymore); (2) the mass of support structures, which need to counteract torques that bend the mirrors, scales with the product of the mass of the optical surface they need to carry and its radius; and (3) about 30\% of the mass of the \textsc{Herschel} telescope were contributed by the optical surfaces and about 70\% by the support structures, the mass of a \textsc{Capella} telescope with a geometric collecting area of 3~m$^2$ is approximately 70~kg.

As an example, Figure~\ref{fig:mirror} shows the design of a segmented SiC primary mirror composed of seven elements. Each element is a regular hexagon with a flat-to-flat diameter of 70~cm, with six elements being arranged side-by-side around a central element; the complete configuration forms an aspheric mirror. The alignment of the elements is accurate to within 1~$\mu$m. Light is reflected from the primary to a circular secondary mirror, and from there through a hole in the central element of the primary toward the Cassegrain focus. The geometric collecting area is 3~m$^2$; the effective collecting area, assuming a total aperture efficiency of 80\%, is $A_{\rm e} = 2.4$~m$^2$.

Given that \textsc{Mimosa} is intended to serve as a pathfinder mission, we foresee the use of a monolithic parabolic SiC primary mirror with an aperture of 1~m and thus a geometric collecting area of 0.75~m$^2$; the effective collecting area, assuming a total aperture efficiency of 80\%, is $A_{\rm e} = 0.6$~m$^2$. Using again the \textsc{Herschel} primary mirror as benchmark, scaling down to the dimensions of the \textsc{Mimosa} telescopes leads to a mass of approximately 13~kg.

\subsection{Receiver \label{sec:receiver}}

\textsc{Mimosa} is intended to observe at frequencies around 100~GHz; we foresee the use of one single-circular polarization, single-side band (SSB) high-electron-mobility transistor (HEMT; see., e.g., Chapters 5.2.4 and 5.2.5 of \citealt{wilson2009}) receiver per satellite. Using heterodyne HEMT receivers for VLBI is a mature technology (e.g., \citealt{han2013}). A typical receiver combines two HEMT amplifiers, a band-defining filter, a mixer -- which combines the sky signal with a reference wave from a local oscillator (LO) and maps it into the intermediate frequency (IF) band --, and a low-noise amplifier. Outside the receiver, we require a tunable LO and the IF circuit. In order to balance (continuum) sensitivity and data rates, we plan with an instantaneous bandwidth of 4096~MHz. We expect a receiver noise temperature of 40~K. The receiver and the IF circuit require a total power of 10~W; including the receiver optics, their total mass is 10~kg. We assume that the LO unit requires another 30~W of power and comes with a mass of 5~kg.

HEMT receivers require a stable operating temperature of about 20~K. Commercially available space-qualified two-stage cryocoolers (like the \href{https://www.shi.co.jp/industrial/en/product/science/space/20k-stirling-cooler.html}{Sumitomo LB2ST}) provide a cooling power of 200~mW at the 20-K stage, have a mass of 10~kg, and consume 90~W of power. We conservatively assume a cryostat mass of 20~kg.

For the \textsc{Capella} observing frequencies around 690~GHz, ample experience with the design of heterodyne receivers is available, especially thanks to the development and serial production of the Band~9 cartridge-type receivers for the Atacama Large Millimeter Array (ALMA) \citep{baryshev2015}. We aim at a simple design that complies with the restrictions with respect to mass, volume, and power consumption associated with a space telescope. We foresee a heterodyne double-sideband (DSB), single-circular polarization receiver using a niobium superconductor-insulator-superconductor (SIS) mixer with an instantaneous bandwidth of 4096~MHz. Since it is important to keep the receiver noise temperature (which, for a space telescope, is the only relevant contribution to the system temperature) low, we conservatively restrict the radio frequency range to the interval 680--700~GHz (corresponding to 8700~km\,s$^{-1}$ in velocity) for the remainder of this whitepaper, thus allowing for optimization of the noise temperature for a small frequency range. We expect DSB noise temperatures $\lesssim$100~K across the range of observing frequencies, which is about three times the quantum limit of 33~K at 690~GHz.

The SIS mixer needs to be supplied with a reference wave from a local oscillator (LO). The \textsc{Herschel} HIFI instrument \citep{degraauw2010} used a combination of an yttrium iron garnet (YIG) oscillator and an amplifier multiplier chain; using this design as a reference, we may expect that the LO unit requires a power of about 20~W and has a mass of 5~kg.

Niobium SIS mixers require operating temperatures of around 4~K. This requires placing the receiver into a staged cryocooler. For astronomical space missions, a dedicated 4-K Joule-Thomson cryocooler, which combines a 20-K two-stage Stirling precooler with a 4-K Joule-Thomson stage with 40~mW cooling power, is available \citep{sato2021}. The total mass of the cryocooler system is 40~kg, its maximum drive power is 180~W; it is designed for a life time of at least three years.

The ALMA Band 9 cartridge receiver, without a cryostat, has a mass of about 10~kg (A. Gonzalez, private communication). The receiver cartridge and the cold head can be placed together inside a cryostat with a volume of 22~liters \citep{sekimoto2003}. For \textsc{Capella}, receiver components inside the cryostat include the feedhorn, a circular polarizer, the SIS mixer, and a 30dB cryogenic amplifier; whereas the feedhorn and the polarizer are passive elements, the mixer and the cryogenic amplifier consume up to about 5~W and 0.1~W of power, respectively. Outside the cryostat, two additional 30dB low-noise amplifiers will be needed; each will consume about 10~W.

Adding up the numbers for all components, the receiver system (receiver, LO, and coolers) will require a power supply of about 225~W. Assuming a receiver mass of 10~kg and a cryostat mass of 20~kg, the total mass of the receiver system will be about 75~kg.

\subsection{Sampler and Recorder \label{sec:recorder}}

In order to transmit the data to a ground station and to accommodate time delays between reception and transmission of data, the signal from the receiver needs to be digitized and recorded. As a compromise between data quality and data rate, modern VLBI digital backend systems use two-bit (four level) quantization; current maximum data rates are 64~Gbps, corresponding to a Nyquist-sampled signal with a bandwidth of 16~GHz \citep[e.g.,][]{vertat2015}. For \textsc{Capella} and \textsc{Mimosa}, a bandwidth of 4096~MHz requires sampling at a rate of 8192 megasamples per seconds; such (single-channel) sampling speeds are available in commercially produced digitizer boards. Commercial off-the-shelf, radiation-hardened, space-rated digitizer boards operating at such sampling rates consume around 5~W of power (e.g., the Texas Instruments \href{https://www.ti.com/product/ADC12DJ5200-SP}{ADC12DJ5200-SP}) and have masses of a few hundred grams at most; in our case, we conservatively assume a power consumption of 10~W and a total package mass of 2~kg.

The recording system needs to accommodate the high data rates of 16384~Mbps (16~Gbps, assuming standard 2-bit sampling) and large data volumes. Commercially produced solid state drives (SSDs) can store up to 100~TB of data (\href{https://nimbusdata.com/products/exadrive/specifications/}{Nimbus Data ExaDrive DC}); internal (connected through PCI Express 4.0 or faster) drives achieve read/write speeds up to around 120~Gbps (e.g., \href{https://www.gigabyte.com/kr/SSD/AORUS-Gen4-AIC-SSD-8TB#kf}{AORUS Gen4 AIC SSD 8TB}). High-end SSDs consume approximately 1~W/TB of power when active and about 0.1~W when idle; their endurance is around 1000 times their storage capacity (in order to achieve high endurance, single-level cell SSDs are strongly preferred). At a data rate of 16~Gbps, a disk pack with 64 8-TB SSDs can store a total of 250,000 seconds of data. Assuming conservatively that at any given time two SSDs (one storing data, one being read out) are active results in a total power demand of about 23~W. Operating hard drives alternately leads to an endurance of the disk pack of at least four years. Requiring that the disk pack is sufficiently shielded or hardened against radiation, we assume its mass to be 5~kg.

\subsection{Frequency Standard \label{sec:oscillator}}

Accurately preserving the phase of the light received from astrophysical sources requires a highly accurate frequency standard that supplies the receiver and the sampler with stable reference frequencies. The degree of correlation $C$ of the signals from two telescopes depends on the integration time $\tau$, the observing frequency $\nu$, and the (integration time dependent) Allan deviation (i.e., the square root of the Allan variance) of the frequency standard $\sigma_{\rm y}(\tau)$ like
\begin{equation}
\label{eq:coherence}
C(\tau,\nu,\sigma_{\rm y}) = \exp\left[-(2 \pi \nu \tau \sigma_{\rm y})^2\right]
\end{equation}
\citep[see, e.g., Chapter 9.5.2 of][]{thompson2017}. The type of frequency standard and the combination of integration time, observing frequency, and Allan variance must be chosen such that $C$ remains reasonably close to unity. In our case, this choice is complicated by the high observing frequency and the tight restrictions on mass and power consumption of the frequency reference.

VLBI networks almost universally use active hydrogen masers; however, even recent space-qualified designs come with uncomfortably high mass and power consumption figures \citep[see, e.g.,][]{caccia2011}. A better choice is offered by highly stable quartz crystal oscillators which recently have reached stability levels similar to those of space-qualified active hydrogen masers. The AccuBeat Ultra Stable Oscillator (\href{https://www.accubeat.com/uso}{USO}), which was designed for the ESA \href{https://www.esa.int/Science_Exploration/Space_Science/Juice}{JUICE} mission to Jupiter, provides an Allan deviation of $\sigma_{\rm y}(\tau) = 1\times10^{-13}$ for $\tau$ in the range from one second to several minutes. Inserting this value into Equation~\ref{eq:coherence} and assuming $\tau = 5$~s and $\nu = 100$~GHz (for \textsc{Mimosa}) results in $C = 0.91$. Likewise, using $\tau = 1$~s and $\nu = 690$~GHz (for \textsc{Capella}) results in $C = 0.83$. Those degrees of coherence are acceptable, although they imply rather short integration times. The USO has a nominal power demand of 7~W and a mass of 2~kg \citep{accubeat2023}.

For completeness, we note that various alternative options for frequency standards are either already available or under active development. The off-the-shelf Vector Atomic \href{https://vectoratomic.com/eg30-advanced-release}{Evergreen-30} rackmount optical clock offers $\sigma_{\rm y}(\tau=1{\rm\,s}) \approx 2.5\times10^{-14}$ at a mass of 20~kg and a power consumption of 80~W. Recent progress in the development of thorium nuclear optical clocks \citep{zhang2024} promises the development of solid-state atomic clocks that are substantially smaller than existing atomic clocks and, thanks to their mechanical robustness, would be ideally suited for space applications.

\subsection{Satellite Positioning and Clock Synchronization \label{sec:gnss}}

An instantaneous bandwidth of $\Delta\nu = 4096$~MHz (Section~\ref{sec:receiver}) implies a coherence time of $\tau_{\rm c} \approx 1/\Delta\nu \approx 0.24$~ns and thus a coherence length of $c\tau_{\rm c} \approx 7$~cm (with $c$ being the speed of light); if needed and appropriate, longer coherence times and lengths can be achieved by subdividing the signal into narrower bands (to be discussed in Section~\ref{sec:correlation}). When data from two observatories are correlated, the optical path difference needs to be controlled and compensated to within a small fraction of the coherence length. To facilitate a straightforward data correlation, the relative position of the telescopes should be known with an accuracy better than the coherence length at the time of correlation (not necessarily at the time of observation). This was not possible for observatories on high Earth orbits, for which positions could be determined only within tens to hundreds of meters \citep[e.g.,][]{zakhvatkin2020}.

As noted by, e.g., \citealt{kudriashov2021}, global navigation satellite systems (GNSS), and especially the Global Positioning System (GPS), can provide accurate position information for satellites on low Earth orbits. Commercially produced GPS receivers for LEO satellites regularly provide absolute post-processing position accuracies of few centimeters and velocity accuracies of few millimeters per second (e.g., \href{https://www.moog.com/content/dam/moog/literature/sdg/space/avionics/Moog-TriGROTriGPOD-Datasheet.pdf}{JPL BlackJack+TOGA}) which is well within our requirements. Masses of such systems are around 3~kg, the power consumption is around 20~W. Reduced-dynamic, post-processing precise orbit determination, which combines GNSS position measurements with dynamic modelling of the satellite orbit, regularly achieves ($1\sigma$) accuracies $<$5~cm \citep{allah2022}.

Since each satellite will use its own on-board frequency reference standard, GNSS receivers are also required for synchronizing the individual clocks to a common global reference time and accurately time-stamping the recorded data, by means of a shared one-pulse-per-second (1PPS) reference signal. This is possible to accuracies on the order of nanoseconds to tens of nanoseconds (depending on the equipment and the integration times used), and is a standard procedure in VLBI.

\subsection{Data Downlink \label{sec:downlink}}

In order to allow for a reasonable duty cycle of \textsc{Mimosa} and, especially, \textsc{Capella}, the high data rates of 16~Gbps in each satellite need to be matched by sufficiently high downlink speeds. Such downlink speeds from LEO to the ground can be achieved with near-infrared laser communication systems \citep[e.g.,][]{carrasco2017}. The NASA TeraByte InfraRed Delivery system demonstrated data rates up to 200~Gbps (2 channels $\times$ 100~Gbps) from a cubesat \citep{schieler2023}. Various commercial developers (e.g., \href{https://www.spacebeam.co.kr/}{SpaceBeam Inc.}, which recently established the Osong Optical Ground Station) are currently aiming at further improving optical free space communication systems. System masses are around 5~kg, the amount of power required is around 50~W.

From a given point on the surface of Earth, a LEO satellite at an altitude of 450~km (600~km) is visible for about 11\% (13\%) of its orbital period (i.e., about 10 to 12 min out of 95 min) at most. Accordingly, a global network of several optical ground stations should be provided in order to avoid substantial dead times in the operation of the VLBI array. Since we assume polar orbits, sites at high northern or southern latitudes are preferred; example candidate sites are Dasan Station, Ny-Alesund, Spitsbergen (located at 78$^{\circ}$55$'$ North) and Jang Bogo Station, Terra Nova Bay, Antarctica (located at 74$^{\circ}$37$'$ South). From the latitude of Jang Bogo Station, a satellite on a polar orbit at an altitude of 450~km is visible at every orbit at a minimum transit elevation angle of about 6$^{\circ}$; at higher latitudes and/or for higher orbits, the minimum transit elevation angle is larger.

\subsection{Pointing Control \label{sec:pointing}}

Assuming that each \textsc{Capella} telescope forms a non-circular aperture with a maximum diameter of $d = 2.1$~m and observes at a wavelength $\lambda = 434~\mu$m, the minimum resolution angle of a single telescope is $\phi \approx 1.2\lambda/d \approx 51''$; for \textsc{Mimosa} with $d=1$~m and $\lambda=3$~mm, we have $\phi \approx 12'$. In order to avoid significant losses, the telescope -- actually, the entire satellite -- needs to be pointed at its target with an accuracy corresponding to a small fraction of the beam size, meaning here a few arcseconds.

For small satellites, the necessary accuracies can be achieved by a combination of a star tracker and control moment gyroscopes (CMGs). Star trackers combine a wide-angle optics system and a CCD camera to track the positions of bright stars and thus to measure the orientation of the spacecraft. A modern, commercially produced star tracker (e.g., Jena-Optronik \href{https://www.jena-optronik.de/products/star-sensors/astro-aps.html}{ASTRO APS3}) has a r.m.s. pointing accuracy of about $1''$, a mass of about 2~kg, and consumes up to about 8~W of power. Since star trackers need to avoid pointing both at the Sun and the Earth, we conservatively assume that three star trackers, pointing into different directions, need to be attached to each satellite, with one of them being active at any given time.

CMGs store angular momentum which can be transferred to the spacecraft in order to rotate it. Full control of all three axes requires three CMGs; for redundancy, usually four elements, arranged in a pyramidal geometry, are installed. A modern four-element CMG system for small satellites (e.g., 4$\times$\href{https://tensortech.co/product/detail/cmg_for_over_30_kg}{TensorCMG-1}) has a mass of about 10~kg and a power consumption of 20~W typically and about 40~W at most. The pointing accuracy of the satellite is usually limited by the accuracy of the star tracker.

\subsection{Power and Mass Budgets \label{sec:budgets}}

\begin{table}[t!]
\centering
\setlength{\tabcolsep}{30pt}  
\begin{tabular}{lrr}
\toprule 
Component & Power (Watt) & Mass (kg) \\ 
\midrule 
Telescope & 0 & 70 \\
Receiver (including cryostat) & 25 & 30 \\
Local oscillator unit & 20 & 5 \\ 
Cryocooler & 180 & 40 \\ 
Sampler & 10  & 2 \\
Recorder & 23 & 5 \\
Crystal oscillator & 7 & 2 \\ 
GNSS receiver & 20 & 3 \\
Laser downlink & 50 & 5 \\
Star trackers & 8 & 6 \\
Control moment gyroscopes & 20 & 10 \\
Microcomputers & 20 & 8 \\ 
\midrule
Total & 383 & 186 \\ 
\bottomrule   
\end{tabular}
\caption{Power and mass budgets for the science payload of a \textsc{Capella} satellite. We assume three star trackers with a total mass of $3\times2$~kg, with one of them being active and consuming power at any given time ($1\times8$~W). \label{tab:capella}}
\end{table}

\begin{table}[t!]
\centering
\setlength{\tabcolsep}{30pt}  
\begin{tabular}{lrr}
\toprule 
Component & Power (Watt) & Mass (kg) \\ 
\midrule 
Telescope & 0 & 13 \\
Receiver (including cryostat) & 20 & 30 \\
Local oscillator unit & 30 & 5 \\ 
Cryocooler & 90 & 10 \\ 
Sampler & 10  & 2 \\
Recorder & 23 & 5 \\
Crystal oscillator & 7 & 2 \\ 
GNSS receiver & 20 & 3 \\
Laser downlink & 50 & 5 \\
Star trackers & 8 & 6 \\
Control moment gyroscopes & 20 & 10 \\
Microcomputers & 20 & 8 \\ 
\midrule
Total & 298 & 99 \\ 
\bottomrule   
\end{tabular}
\caption{Same as Table~\ref{tab:capella} but for \textsc{Mimosa}. \label{tab:mimosa}}
\end{table}

The discussion provided in the previous subsections makes it possible to present the power and mass budgets for the science payloads of our VLBI satellites. Whereas these budgets are necessarily approximations, they are based on existing, and partially commercially produced, technology. We summarize the power and mass budgets in Table~\ref{tab:capella} for \textsc{Capella} and in Table~\ref{tab:mimosa} for \textsc{Mimosa}, respectively. In addition to the individual contributions discussed in Sections \ref{sec:telescope} through \ref{sec:pointing}, we take into account that each of the eight active components will need control electronics. Using a low-key (quad core 1.2~GHz 64-bit CPU, 1~GB RAM) microcomputer like the \href{https://www.raspberrypi.com/products/raspberry-pi-3-model-b/}{Raspberry Pi 3 Model B} as benchmark, we may expect a \href{https://www.pidramble.com/wiki/benchmarks/power-consumption}{typical} power consumption around 2.5~W. Assuming a mass of 1~kg including a protective casing, we arrive at a total mass of 8~kg and a total power consumption of 20~W for eight units.

Adding up all individual contributions, we find for the payload of \textsc{Capella} a power consumption of 383~W and a mass of 186~kg, and for the one of \textsc{Mimosa} a power consumption of 298~W and a mass of 99~kg. We note that some of the power values we quote are nominal values; temporary peak or low values may differ substantially from the nominal values, and may require adjustment of satellite operations to keep the total power demand within specifications. We further note that, technically, the pointing control system (star trackers and CMGs) is not a part of the science payload but of the satellite main bus; we nevertheless added it to our budgets since it is essential for the science operations. Given the payload masses, we assume a total launch mass of approximately 500~kg per \textsc{Capella} satellite and of approximately 250~kg per \textsc{Mimosa} satellite.

\section{Sensitivity \label{sec:sensitivity}}

The ability of \textsc{Capella} to address various science cases is determined not only by its angular resolution but also by its sensitivity. The radiometric formula provides the $1\sigma$ sensitivity for a network of $N$ antennas operating for an on-source observing time $\Delta t$ with an instantaneous bandwidth $\Delta\nu$, antenna efficiency $\rho_{\rm A}$, and system temperature $T_{\rm sys}$,
\begin{equation}
\label{eq:radiometry}
\sigma = \frac{\rho_{\rm A} T_{\rm sys}}{0.88\,C \sqrt{N(N-1) \Delta\nu \Delta t}} ~ ;
\end{equation}
here, $C$ denotes the degree of coherence determined by Equation~\ref{eq:coherence}; in the following, we will assume $\tau=1$~s and $C=0.83$ for \textsc{Capella}, and $\tau=5$~s and $C=0.91$ for \textsc{Mimosa} (see Section~\ref{sec:oscillator}). The additional factor 0.88 results from signal quantization losses for 2-bit (4-level) sampling (cf. Section~\ref{sec:recorder}).

Various sensitivity estimates result from different assumptions. The \emph{total point source sensitivity} assumes that all baselines are observing the same point source all the time; in this case, $N = 4$ for \textsc{Capella} and $N=2$ for \textsc{Mimosa}. For $\Delta t = 250$,000~s and the parameter values given in Section~\ref{sec:interferometer}, one finds $\sigma \approx 1.4$~mJy for \textsc{Capella} and $\sigma \approx 5.1$~mJy for \textsc{Mimosa}. Since a given satellite will usually be shadowed by Earth for parts of each orbit, the total observing time to be allocated will be longer than the actual on-source time. The total point source sensitivity describes the noise level of the final map of the target and is the fundamental sensitivity limit.

The other extreme of sensitivity estimates is given by the \emph{instantaneous sensitivity per baseline} which assumes that each baseline needs to be able to detect the target within an observing time short enough to ensure that the given baseline only moves through a small part of the $uv$ plane; this condition may apply to targets with very complex and/or rapidly variable spatial structure, or to situations where a nearly constant visibility is required for calibration purposes (especially fringe fitting). The maximum relative speed of two satellites (for the case of two co-planar circular orbits at altitudes of 450~km and 550~km, respectively) is about 15.2~km\,s$^{-1}$. In this case, the length of the baseline changes by 11\% of the orbit diameter within about 100 seconds. Inserting thus $\Delta t = 100$~s and $N=2$ (and the other parameter values from Section~\ref{sec:interferometer}) into Equation~\ref{eq:radiometry}, one finds $\sigma \approx 0.17$~Jy for \textsc{Capella} and $\sigma \approx 0.25$~Jy for \textsc{Mimosa}. As noted at the beginning of this section, we assume coherence times of at most five seconds; integration times beyond this limit will require coherent averaging, potentially using iterative correlation schemes, over multiple time segments. Alternatively, \emph{incoherent} averaging over multiple time segments may be used. There, the flux detection threshold follows from multiplying Equation~\ref{eq:radiometry}, applied to a single time segment (1~s for \textsc{Capella} and 5~s for \textsc{Mimosa}), with a factor that itself depends on the false alarm probability used as detection limit; for a false alarm probability of 0.01\%, corresponding to a Gaussian significance of $3.9\sigma$, this factor is  $2.06\,N'^{-1/4}$, with $N'$ being the number of time segments \citep{rogers1995}. For \textsc{Capella}, with $\Delta t = 1$~s and $N'=100$, the detection threshold thus defined is about 1.1~Jy; this may be compared to the value of 0.68~Jy ($3.9\sigma$) expected from coherent averaging over 100 seconds. In case of a global search across all baselines, these values are reduced by a factor $\sqrt{2/(N(N-1))}$, resulting in 0.46~Jy and 0.28~Jy for $N=4$, respectively. For \textsc{Mimosa}, with $\Delta t = 5$~s and $N'=20$, the detection threshold thus defined is about 1.1~Jy; this may be compared to the value of 0.99~Jy ($3.9\sigma$) for coherent averaging over 100 seconds.

In addition to sensitivity considerations, optimizing the image fidelity implies a characteristic binning time scale for the $uv$ plane data by demanding that beam smearing be minimized. A crude estimate of this time scale follows from the condition that radial and circumferential smearing should be about equal; this implies (cf. Equation~10.100 of \citealt{thompson2017}) $\Delta\nu / \nu \approx \omega\Delta t$, with $\omega$ being the angular speed of the observatory around Earth and $\Delta t$ being the integration time. For \textsc{Capella}, $\Delta\nu / \nu = 4/690 = 5.8\times10^{-3}$ and $\omega \approx 1.12\times10^{-3}$ (i.e. the angular speed of a satellite orbit at an altitude of 450~km), and thus $\Delta t \approx 5.2$~s. For \textsc{Mimosa}, $\Delta\nu / \nu = 4/100 = 0.04$ and $\omega \approx 1.10\times10^{-3}$ (i.e. the angular speed of a satellite orbiting at an altitude of 600~km), and thus $\Delta t \approx 37$~s. (Those values may be compared to the corresponding time scale for ground based observations, which is on the order of tens to hundreds of seconds).

\section{Data Postprocessing \label{sec:postprocessing}}

\subsection{Data Formatting \label{sec:formats}}

The data from the receiver are digitized, time-tagged, and written to a hard disk. The high sampling rates of VLBI, as well as the required timing accuracies, require a dedicated data format; such a format is provided by the VLBI Data Interchange Format (VDIF; \citealt{whitney2010}). In VDIF, data are arranged into a stream of Data Frames, each containing a short self-identifying Data Frame Header, followed by a Data Array which contains the actual samples. Data Frame Headers are either 16 or 32 bytes in length and contain, among other parameters, a time stamp in the format ``seconds from reference epoch / data frame length in bytes / data frame number in given second''. Data Frames are limited to sizes between $2^6$ and $2^{27}$ bytes. In general, downlink utilization is better with large frames. Existing VLBI backend designs output data onto Ethernet, hence the VDIF frame payload is commonly between 1024 and 8192 bytes (sometimes 32,768 bytes) in size. Notably, VDIF supports any number of bits per sample from 1 to 32; this will be relevant in Section~\ref{sec:correlation}.

\subsection{Corrections for Orbital Kinematics \label{sec:corrections}}

In the course of its orbit around Earth, the speed of a satellite relative to the target changes by up to $\pm$7.6~km\,s$^{-1}$ if the orbital plane is parallel to the line of sight. The resulting frequency shifts (by up to 18~MHz for \textsc{Capella} and 2.5~MHz for \textsc{Mimosa}) need to be compensated (1) by utilizing active Doppler correction, where the frequency of the local oscillator is continuously modified such that the observed frequency of light from the target does not change; and/or (2) adjusting the observed frequency at the correlation stage.
%Since we foresee a channelization of data before correlation (to be discussed in Section~\ref{sec:correlation}), active Doppler correction (option 1) will be necessary to prevent shifting data out of their spectral bands.
The latter option requires the construction of accurate orbit models that can be applied to the receiver settings in real time.

Relativistic effects need to be considered, and corrected for, in addition. A speed of 7.6~km\,s$^{-1}$ relative to the target and/or the correlator corresponds to a Lorentz factor $\gamma = 1 + 3.2\times10^{-10}$, resulting in a special relativistic Doppler shift of 220~Hz at 690~GHz and 32~Hz at 100~GHz, and a time dilation of 0.32~ns per second. Since our VLBI satellites move at different distances from the center of Earth, they also experience differential gravitational time dilations and differential gravitational redshifts; both effects scale with a factor
\begin{equation}
\label{eq:gravredshift}
\Lambda = \frac{G M_\oplus }{c^2} \left( \frac{1}{R_1} - \frac{1}{R_2} \right)
\end{equation}
(with $G$ being Newton's constant, $M_\oplus$ being the mass of Earth, and $R_{1,2}$ being the two orbit radii) to very good approximation for our range of orbit radii. For orbit altitudes of 450~km and 600~km, respectively, we have $\Lambda = 1.4\times10^{-11}$; this corresponds to a differential redshift of 9.7~Hz at 690~GHz and a differential time dilation of 14~ps per second.

\subsection{Correlation \label{sec:correlation}}

For VLBI observations, Fourier transform (FX) correlation, where the cross-correlation of two signals (which are functions of time) is computed by multiplying their Fourier transforms (which are functions of frequency, i.e. spectra), thus exploiting the cross-correlation theorem, is the method of choice. The public Distributed FX-architecture (DiFX) software correlator \citep{deller2011} is nearly universally used for this purpose; a DiFX branch dedicated to networks including the RadioAstron space telescope, ra-DiFX, is available \citep{bruni2016}. DiFX correlates the data from each pair of telescopes while applying the geometric delay (i.e. the difference in optical path length along the line of sight) between the telescopes, the change of the delay with time (delay rate), and, in case of ra-DiFX, changes of the delay rate with time (delay acceleration).

Since the positions and velocities of the \textsc{Capella} satellites will be known with accuracies of few cm and few mm\,s$^{-1}$, respectively, at any time (cf. Section~\ref{sec:gnss}), the values of the delays, delay rates, and delay accelerations will be known a priori, at least to good approximation. For an instantaneous bandwidth of 4096~MHz, the coherence time is about 0.24~ns, the corresponding coherence length is about 7~cm. Coherence time and length can be enlarged by subdividing the full bandwidth into multiple channels and correlating channel by channel. In case of a channelization into a $16\times256$~MHz set of bands, the coherence times are about 4~ns, the coherence lengths are about 1.2~m; in turn, the sensitivity thresholds (Section~\ref{sec:sensitivity}) increase by a factor of 4. Such coherence lengths are, at least in theory, large enough to correlate data ``blindly'' using a theoretical correlator model and to populate the $uv$ plane with visibilities, without the need for fringe searches. The channelization can be achieved by passing the recorded data through a digital polyphase filter bank; to keep additional quantization losses negligible ($\lesssim$1\%), the output of the filter bank should be sampled at four bits (16 levels) per sample or more. We note that previous space-VLBI missions were known to require elaborate fringe search techniques, for which ample experience from the RadioAstron mission is available \citep{gomez2016,kim2023,savolainen2023}.

\subsection{Computing Resources \label{sec:compute}}

We envision the use of a small computing cluster with normal CPU compute nodes. (To date, development of DiFX GPU acceleration is still in progress.) Judging from the computing and file input/output performance of the correlation of ground based experiments (mm-VLBI, geodesy), a cluster with 1024 CPU cores and a fast cluster interconnect, like 56 Gbps InfiniBand, would be necessary for processing locally stored data at real-time equivalent rates. Data storage requirements for four satellites observing at 50\% duty cycle for one year at 4096~MHz bandwidth and with 2-bit sampling, plus various overheads and the storage of the visibility data, amount to around 130 petabytes per year.

The DiFX software correlator scales well with the number of compute nodes assigned to it, meaning that VLBI scans may be worked through sequentially on the full cluster about as quickly as separate scans run simultaneously on subsets of the cluster nodes. Throughput can be increased by utilizing a larger computer cluster. Postprocessing programs to fringe fit correlator output -- if actually required -- scale less well. Here, ``parallelization'' and a better utilization of cluster computing capacity are possible largely by carrying out the fringe fit of correlator output from different scans at the same time on different compute nodes.

\section{Discussion \label{sec:discussion}}

Multiple cutting-edge science cases require observations at high radio frequencies with a combination of high angular resolution and good image quality, thus requiring the use of very long baseline interferometry with good $uv$ plane coverage. In Section~\ref{sec:science}, we identified photon rings around supermassive black holes, the formation of AGN jets, molecular gas accretion flows in galactic centers, binary supermassive black holes, stellar flares, and stellar novae; naturally, this list is incomplete and cannot account for surprise discoveries that may be enabled by any new observing technique.

We present a tentative design of a space-only VLBI program that includes two missions: \textsc{Mimosa}, a pathfinder mission comprising two small satellites on separate low-Earth orbits and observing at frequencies around 100~GHz, and \textsc{Capella}, a science-grade VLBI network composed of four small satellites on separate low-Earth orbits and observing at frequencies around 690~GHz. (We refer to the two missions collectively as the \textsc{Capella} Program.)

Observing a given target with \textsc{Capella} over several days results in a near-complete $uv$ plane coverage and an angular resolution of about 7~$\mu$as. We emphasize that \emph{only} space-only VLBI networks are able to achieve such a performance given the restrictions on geographic locations of antenna, observing times, and observing frequencies that affect VLBI networks based partially or entirely on the ground. A \textsc{Capella} type network can offer effectively unlimited observing times and can handle target-of-opportunity (ToO) situations within a few hours to a few days after the initial alert. \textsc{Capella} provides observations at very high angular resolution but with limited sensitivity (target flux densities $\gtrsim$10~mJy); this makes it complementary to ground-based observatories, especially \href{https://www.eso.org/sci.html}{ALMA}, that are able to provide high-sensitivity ($\mu$Jy level) observations at limited angular resolutions (tens of mas at best).

Given that there currently is no experience with space-to-space VLBI, we designed \textsc{Mimosa} as a pathfinder mission that is able to test all relevant technologies and methods and to provide a proof-of-concept. Each space-VLBI satellite needs to carry a complete VLBI station, including telescope, receiver, digitizer, recorder, and a frequency standard; has to be able to determine its position within a few centimeters; needs to be able to accurately point at its targets; and has to transmit the recorded data to a ground station with sufficiently high speed. Using data channelization (Section~\ref{sec:correlation}) implies that the receivers should utilize active Doppler correction. The total power demand and mass of the science payload need to be sufficiently small to be integrated into $\leq$500-kg class satellites. Indeed, as discussed throughout Section~\ref{sec:system}, the technology required for each of the components listed above is already available within the desired mass and power limits and, in some cases, is available off-the-shelf (like high-precision GNSS receivers or star trackers). We note that both \textsc{Mimosa} and \textsc{Capella} provide opportunities for the development of novel technologies with improved specifications, given that essentially all of the technologies discussed in Section~\ref{sec:system} have been developed for purposes other than space-VLBI. Since we require several identical VLBI stations for each mission concept, development efforts can be limited to a single satellite design which then is used as a template for a mini-series of identical space telescopes. The relatively limited development efforts, as well as the relatively limited cost of launching up to four small satellites into LEO, suggests that any major economy with access to a space program (like the Republic of Korea) will easily be able to afford the \textsc{Capella} Program economically.

%%% --- Continue here ---

In this whitepaper, we presented a specific ``minimalist'' design for a high-frequency space-only VLBI program. Obviously, many variations of the theme are possible. Increasing the number of satellites -- for either observatory -- increases the sensitivity and the number of baselines, and thus shortens the observing time needed to meet specific sensitivity or $uv$ plane coverage requirements. The (currently) guaranteed life times of critical components, especially the receiver cooling system (cf. Section~\ref{sec:receiver}) imply that (currently) the minimum mission life time of any given \textsc{Capella} satellite is limited to about three years; accordingly, one might consider gradual replacement of aging satellites with updated ones, meaning a gradual upgrade of the capabilities of the constellation as a whole.

Our choice of observing frequencies around 100~GHz for \textsc{Mimosa} reflects a balance between technical simplicity (HEMT receivers instead of SIS mixers) and proximity to the frequency range of scientific interest. Considering substantially lower frequencies might simplify the design, and accelerate the timeline for construction and deployment, further. The observing frequencies around 690~GHz for \textsc{Capella} are motivated by a combination of high angular resolution, decent sensitivity, building on existing technology, and being able to complement ground-based observatories (especially ALMA); both higher (e.g. for improved angular resolution) or lower (e.g. for improved sensitivity, or complementarity with other missions like BHEX) frequencies might be chosen instead. In order to keep the system simple, we restricted ourselves to a single-frequency band single-polarization receiving system; if power, mass, and data rate limits permit, upgrading the receiver capabilities to dual polarization and/or multiband observations is straightforward and would be desirable. 

Whereas this whitepaper has a strong focus on the satellite constellations, actually utilizing their capabilities requires a dedicated infrastructure on the ground. As noted in Section~\ref{sec:downlink}, several ground stations for laser communication will be necessary. Once data have been received and recorded at a ground station, they need to be transmitted or physically transported to a correlator. The correlator combines the signals from each pair of telescopes coherently and derives interferometric visibilities as function of the coordinates $(u,v)$; dedicated FX software correlators (DiFX, ra-DiFX) are already available and can be adapted to the needs of \textsc{Capella}. Even if the capabilities of each satellite are strongly constrained by mass, power, data rate, and coherence time limits, a combination of long (several days) observing times and sophisticated postprocessing schemes will make \textsc{Capella} a powerful observatory -- the first of its kind.

\section{Summary and Conclusions \label{sec:summary}}

Motivated by the potential to explore various astrophysical phenomena on as-yet unprobed spatial scales, we designed the \textsc{Capella} Program which ultimately aims at the deployment of a space-only VLBI system composed of four small ($\leq$500~kg) satellites on low Earth orbits. Our key results are as follows:
\begin{enumerate}
\item  At an observing frequency of 690~GHz, the \textsc{Capella} observatory is able to achieve an angular resolution of about 7~$\mu$as and a near-complete $uv$ plane coverage when observing a given target over a period of around three days; \textsc{Capella} will thus be unprecedented in terms of angular resolution as well as image quality. Targets located anywhere in the sky and with flux densities $\gtrsim$10~mJy can be observed even on short notice.
\item  In order to obtain the necessary experience with constructing, deploying, and operating a space-to-space VLBI array, we propose \textsc{Mimosa} as a dedicated pathfinder mission. Composed of two satellites in LEO and observing at frequencies around 100~GHz, the observatory will be able to provide a convincing proof-of-concept. With angular resolutions as good as 35~$\mu$as, \textsc{Mimosa} will be able to provide unprecedented new observations of a limited number of selected bright targets, despite being not a science mission.
\item  Each VLBI satellite needs to carry all components needed for a complete VLBI station, namely telescope, receiver, sampler, recorder, frequency reference, GNSS receiver, and downlink system, plus a pointing control system (star tracker and control moment gyroscopes). The technology for all required components is already available, partially off-the-shelf. Assuming reasonable specifications for each component, we estimate a nominal total power demand of around 380~W and a total mass of about 190~kg for the science payload of \textsc{Capella}, and around 300~W and 100~kg for the payload of \textsc{Mimosa}. The relatively limited efforts needed for developing and launching a VLBI constellation will make the \textsc{Capella} Program an inexpensive project (by spaceflight standards). 
\item  We identified photon rings around supermassive black holes, the acceleration and collimation zones of plasma jets emitted from the vicinity of supermassive black holes, the chemical composition of accretion flows into active galactic nuclei through observations of molecular absorption lines, mapping supermassive binary black holes, the magnetic activity of stars, and nova eruptions in symbiotic binary stars as science cases for \textsc{Capella}. Given that \textsc{Capella} will be the first observatory of its kind, surprise discoveries are possible. 
\end{enumerate}

The \textsc{Capella} Program and other, similar projects, are helped greatly by the ``new space age'' with its emphasis on launching multiple small, mass-produced satellites into low Earth orbits with new low-cost launchers. The establishment of a space-only VLBI network is an obvious application of these new capabilities -- and will open a new window for observational astronomy.

\end{document}